\documentclass[11pt,a4paper]{article}
\pdfoutput=1
\usepackage{amsmath,amssymb,amsthm,mathtools,eucal,cite,color}
\usepackage[top=3cm,right=3cm,left=2.5cm,bottom=3cm]{geometry}
\usepackage{framed,graphicx,multirow}
\usepackage[T1]{fontenc} 

\providecommand{\U}[1]{\protect\rule{.1in}{.1in}}
\newtheorem{theorem}{Theorem}

\newtheorem{conjecture}[theorem]{Conjecture}
\newtheorem{corollary}[theorem]{Corollary}

\newtheorem{definition}[theorem]{Definition}

\newtheorem{lemma}[theorem]{Lemma}

\newtheorem{proposition}[theorem]{Proposition}
\newtheorem{remark}[theorem]{Remark}

\def\bpropo#1{\begin{proposition}\label{#1}}
\def\epropo{\end{proposition}}
\def\blemma#1{\begin{lemma}\label{#1}}
\def\elemma{\end{lemma}}
\def\bcor#1{\begin{corollary}\label{#1}}
\def\ecor{\end{corollary}}
\def\bconj#1{\begin{conjecture}\label{#1}}
\def\econj{\end{conjecture}}
\def\brem#1{\begin{remark}\label{#1} \rm}
\def\erem{\end{remark}}
\def\bthm#1{\begin{theorem}\label{#1}}
\def\ethm{\end{theorem}}
\def\bdefi#1{\begin{definition}\label{#1}}
\def\edefi{\end{definition}}

\def\eref#1{(\ref{#1})}         
\def\sref#1{Sect.~\ref{#1}}
\def\aref#1{App.~\ref{#1}}
\def\pref#1{Prop.~\ref{#1}}
\def\lref#1{Lemma~\ref{#1}}
\def\rref#1{Remark~\ref{#1}}

\def\cref#1{Cor.~\ref{#1}}
\def\dref#1{Def.~\ref{#1}}

\def\RR{\mathbb{R}}

\def\HH{\mathcal{H}}

\def\SS{\mathfrak{S}}

\def\inv{^{-1}}

\def\wt{\widetilde}
\def\ol{\overline}
\def\pa{\partial}
\def\lra{\leftrightarrow}
\def\RA{\Rightarrow}

\def\bea#1{
  \begin{eqnarray}\label{#1}}
\def\eea{\end{eqnarray}}
\def\ba{\begin{array}}
\def\ea{\end{array}}
\def\bpm{\begin{pmatrix}} \def\epm{\end{pmatrix}}
\def\ben{\begin{enumerate}}
\def\een{\end{enumerate}}

\def\erw#1{\langle #1\rangle}

\def\wick#1{\,\colon\!#1\!\colon}

\def\ioi{\int_0^\infty}

\numberwithin{equation}{section}
\numberwithin{theorem}{section}

\def\Lp{L^{\rm pt}}

\def\lrpa{\raisebox{1pt}{$\stackrel\leftrightarrow\partial$}}
\def\tree{\big\vert^{\rm tree}}

\def\K{^{\rm K}}
\def\P{^{\rm Pr}}
\def\Vc{V\!\circ\pa}
\def\Vcn{V\!\!\circ\!\pa}

\title{\vskip-10mm How the Higgs potential got its shape}

\author{{\sc Jens Mund$^a$, Karl-Henning Rehren$^b$, Bert
Schroer$^c$} \\ {\small
$^a$ Departamento de F\'isica, Universidade Federal de Juiz de Fora,} \\[-1mm]
{\small Juiz de Fora 36036-900, MG, Brasil, email: jens.mund@ufjf.br} \\[1mm]
{\small $^b$}
  {\small Institut f\"ur Theoretische Physik,
  Universit\"at G\"ottingen,} \\[-1mm] {\small 37077 G\"ottingen,
 Germany,  email: krehren@uni-goettingen.de} \\[1mm]
{\small $^c$ Institut f\"ur Theoretische Physik der FU
  Berlin, 14195 Berlin, Germany,} \\[-1mm] {\small email:
  schroer@zedat.fu-berlin.de}}

\date{}

\begin{document}
\maketitle

\begin{abstract}
String-localized quantum field theory allows renormalizable couplings
involving massive vector bosons, without invoking negative-norm
states and compensating ghosts. We analyze the most general coupling
of a massive vector boson to a scalar field, and find that the scalar
field necessarily comes with a quartic potential which has the precise shape
of the shifted Higgs potential. In other words: the shape of the Higgs potential has not to be
assumed, but arises as a consistency condition among fundamental
principles of QFT: Hilbert space, causality, and covariance. The consistency
can be achieved by relaxing the localization properties of auxiliary quantities,
including interacting charged fields, while observable fields and the
S-matrix are not affected.
This is an instance of the ``$L$-$V$ formalism'' -- a novel model-independent scheme that can be used as a tool to ``renormalize
the non-renormalizable'' by adding a total derivative to the
interaction.
\end{abstract}

\maketitle


\baselineskip14pt

\section{Introduction}
\label{s:intro}

The present study is part of a long-term program \cite{MSY,Sch11,Sch19,GV}
whose aim is to build renormalizable perturbation theory for the Standard
Model on quantum principles (notably Hilbert space and
causality), and detach it from formal quantization based on  classical field
theories which requires to sacrifice the Hilbert space as soon as the
spin or helicity exceeds $\frac12$. Well-established quantitative predictions
are unaltered, but the theoretical reasoning changes from recipes to principles. The latter are powerful enough to 
determine the structure of interactions without invoking classical gauge
symmetry \cite{GMV,GGM}. The program also includes the construction of
``off-shell'' interacting quantum fields. This is a great
  advantage over the BRST method: Recall that the BRST variation of interacting
  charged fields is non-zero (typically a ghost-valued
 gauge transformation). These fields are therefore 
 not defined on the positive
 quotient Hilbert space \cite{KO,Dt}.  In contrast, with the new method, all interacting
 fields are defined on the same Hilbert space. The charged ones,
 however, will have a weaker localization due to the interaction. The
 interaction thus ``selects'' the observables of the theory, namely those
 interacting fields which remain well localized as required by causality. 

The weaker localization properties
of ``off-shell'' (i.e., beyond the S-matrix) charged fields already allowed to re-address and solve
salient infrared problems of QED \cite{MRS2,MRS3}, including the
well-known conflict between locality and the Gauss Law \cite{FPS}, and the
infrared superselection structure \cite{FMS}. We expect that it may
also shed new light on confinement in QCD.

The focus in the present paper is on massive vector bosons.
Massive vector bosons play a central role in the Standard Model,
exhibiting self-couplings and minimal couplings to fermions. 
The immediate problem with these couplings is that the free massive vector 
field on a Hilbert space (the  Proca field with spin 1) has more singular
correlation functions than scalar fields. Its ``short-distance dimension 2''
means that the field causes stronger ultraviolet vacuum polarizations,
which in turn makes ultraviolet divergences in loop diagrams
stronger than with scalar fields. In technical parlance: the interaction density
coupling the massive vector bosons to itself and to Fermi fields (the weak
interaction) is power-counting non-renormalizable.

It has become common practice to make the interaction renormalizable by
using vector potentials on a Krein space, which means that one admits states of
negative norm square. In order to get rid of the latter, one needs
gauge invariance. Gauge invariance not only requires indefinite
metric, it also does not permit a mass term. The Higgs mechanism is
invoked to make the massless gauge bosons behave ``as if'' they were massive  particles. 

The construction to be presented here is an alternative way to secure
renormalizability that
goes without states of negative norm square, ghosts, and spontaneously
broken gauge symmetry. The same  effect of taming the vacuum
polarizations can be achieved by ``allowing more room in space''. This means,
one replaces the Proca fields in the interaction by fields that are localized on
``strings'' (rays extending to infinity), see \sref{s:s-qf}. They live on the Hilbert
space of the massive Proca fields, but have a better UV behaviour. We
show that they can have self-interactions only in the presence of a scalar field,
  and the perturbative
renormalizability of such couplings requires the scalar field to have
a potential in the familiar shape of the shifted Higgs potential
\bea{VH} V(H) \,=\, \frac{m_H^2}2 \big(H^2+  \frac{g}{m}H^3 + 
\frac{g^2}{4m^2}H^4\big) \, =\, \frac{g^2m_H^2}{8m^2}H^2\big(H+\frac{2m}g\big)^2\eea
with its two degenerate minima. 

String-localized quantum field theory (SQFT) offers a variety of tools
to prevent that an interaction involving string-localized fields
makes the entire theory non-local. These tools implement what is
called the ``Principle of String Independence'' (PSI).  The first purpose of this article is to
elaborate the more general ``$L$-$V$ formalism'', of which the PSI is
a prominent instance. We then apply it to the Abelian Higgs Model with only one massive
vector boson. One realizes in
first order of perturbation theory, that the vector boson of mass $m$ must
have a unique cubic coupling to a scalar field called $H$ of
(arbitrary) mass
$m_H$, which may have a cubic self-coupling $H^3$. In second order, the PSI admits a
quartic self-coupling $H^4$. In third order, the PSI uniquely fixes the cubic and quartic
coefficients. Together with the mass term, the outcome
is \eref{VH}.

\eref{VH} is the ``shifted'' version of the symmetric Higgs potential
\bea{VPsi} V(\Psi) = \kappa \Big(\Psi^*\Psi-\frac {v^2}2\Big)^2.\eea
\eref{VPsi} is usually invoked to trigger
the Higgs mechanism, where the complex scalar field $\Psi$ is minimally coupled to a
{\em massless} gauge field. 
The symmetry of \eref{VPsi}
is assumed to be broken spontaneously, and the resulting vacuum
expectation value of the complex field
makes the massless gauge bosons behave as if they were massive 
particles. The physical real Higgs field $H$ describes the
fluctuations of the complex field around its vacuum expectation
value. Expressed in terms of $H$, \eref{VPsi} becomes \eref{VH} with
$m=gv$ and $m_H^2=2\kappa v^2$.

That \eref{VH}
arises as a {\em prediction} of SQFT, rather than an {\em input} to define the
model, is the second main purpose of our article. It retrospectively
justifies the name ``Higgs field'' for the scalar field $H$. But the
mass of the vector boson is not generated by spontaneous symmetry breaking. It is there
from the start.

Popular as it is (and successful as far as the S-matrix is concerned),
we think that the Higgs mechanism suffers from conceptual weaknesses:
To which extent can the degenerate classical minima be regarded as
different ground states of a quantum algebra (which
would justify the term ``spontaneous symmetry breaking'')? The
very (perturbative) construction of such an algebra already picks one of
the classical minima to expand around\footnote{The Feynman rules that
  give the vector boson mass derive from the ``shifted''
  Lagrangian.}. Moreover, the algebra cannot be
constructed on a Hilbert space. The latter has to be recovered with 
the help of compensating ghost fields and the principle of BRST
invariance. The interacting Higgs and other fields of interest are not
BRST invariant and hence are not defined on the BRST Hilbert space.

These detriments can be avoided with SQFT, without compromising on the
fundamental principles of quantum theory.  
Remarkably, the {\em outcome} of the SQFT approach to the Abelian
Higgs Model is equivalent to the {\em input} of the Higgs mechanism:
the presence of a neutral Higgs particle with the potential \eref{VH}
with its two degenerate minima. Gauge symmetry is not assumed 
in the SQFT approach (and not spontaneously broken). 

The Abelian Higgs Model was previously treated in
\cite{GB} and \cite[Sect.~5.1]{DGSV}, see also \cite[Chap.~4.1]{Scha}, 
in the causal BRST setting, in order to equally emphasize the latter fact: that gauge
symmetry needs not to be assumed. In their setting, the shape of the Higgs
potential was inferred from the consistency of the BRST method in
higher orders of perturbation theory. Parts of our analysis are in
fact quite similar to theirs, with the PSI in the place of BRST invariance. But
SQFT goes a step further by working in a Hilbert space and with only physical degrees of freedom from the
outset.

As said before, this is made possible by admitting the vector
potential for the massive particle to be string-localized (see
\sref{s:s-qf}). One can then write down a renormalizable interaction
density and establish that the resulting theory is 
equivalent (in a sense to be explained in \sref{s:PSI-imp}) to the theory with the
non-renormalizable interaction density. We shall show this for the
coupling to the Higgs field in \sref{s:ahm}, and point out that the coupling to Dirac
fields can be added without difficulties, see \sref{s:fermi}.

The Abelian Higgs Model is only a toy model for the weak interaction. The
actual weak interaction (with four vector bosons, electrons
and neutrinos, and Higgs with their experimentally given masses) was
treated without Higgs mechanism in the BRST 
setting \cite{AS,DS,ADS,Scha}, and in SQFT \cite{GMV,GGM}. 
In these papers, it was found by way of necessary consistency
conditions, that the coefficients of cubic self-interaction among several
vector bosons {\em must be} the structure constants of a Lie algebra of
compact type, and in the same way the quartic self-couplings are found. Moreover, the coupling of the vector
bosons to the electrons (neutrinos) {\em must be} chiral (completely chiral), and
Yukawa couplings of the Higgs can only involve scalar Fermi
currents.

The decisive mechanism in the BRST setting (in \cite{ADS,AS,DS,GB,Scha}) is
the consistency of BRST invariance of the S-matrix in higher orders of perturbation theory, as
the necessary tool to recover Hilbert space positivity, while locality
is manifest. The decisive
mechanism in SQFT (in \cite{GMV,GGM} and the present paper) is the
consistent implementation of the PSI in higher orders of perturbation theory, as
the necessary tool to control locality, while positivity is manifest. Neither uses gauge
symmetry. The universality of results (Yang-Mills type of self-couplings,
chirality of couplings to Fermi currents, and the shape of the Higgs
potential) in several variants of BRST and SQFT (see \sref{s:comp})
rather signals an intrinsic consistency between the fundamental
principles of Hilbert space positivity and locality, which shows
up in many different guises.

\paragraph{Plan of the paper.} In \sref{s:sloc}, we briefly recall the
definition and properties of string-localized quantum  fields and the
idea how to use them to improve the renormalizability of perturbative
quantum field theories that are power-counting non-renormalizable.
The S-matrix and interacting quantum fields are constructed
perturbatively along the lines of  ``causal perturbation theory''
\cite{EG,Dt}, which best permits to control the locality of interacting
fields.

We then formulate (in a model-independent way) the PSI to ensure
that the resulting theory does not depend on the auxiliary string
variables, and develop in quite some detail the recursive scheme that generically induces higher order interaction terms. This
``induction'' mechanism is what in the model of our interest eventually produces the 
potential \eref{VH}.

There are actually two variants of the PSI, referred to as ``$L$-$V$'' and
``$L$-$Q$'' formalism, respectively (\sref{s:PSI-imp}). While SQFT
deploys its full power in the former,
the latter is much easier, and therefore best suited to familiarize oneself with the
calculus (cancellation of ``obstructions''). For this reason, we shall
also present in the application to the model (\sref{s:ahmLQ}) the
$L$-$Q$ computations in more detail, mostly referring for the $L$-$V$
variant to ``straightforward computations'' in \sref{s:ahmLV}.
It is worthwhile to mention that the $L$-$V$ formalism is also useful
outside SQFT, whenever one wants to assess the effect of total
derivatives in the interaction, see \sref{s:LVLQ}, and \sref{s:Krein}
for an example.

The core section is \sref{s:ahm}. We apply the method to the Abelian Higgs Model, whose
only input is the free field content: a string-localized massive vector
field (constructed from the Proca field), and a canonical scalar
field (and no ghosts or Stückelberg fields). The PSI in first order determines a unique
cubic coupling among these fields plus a cubic self-coupling of  the
scalar field with an undetermined coefficient. In second order, a quartic self-coupling is induced, and
string independence in third order fixes the cubic and quartic
coefficients. The outcome is the Higgs potential. 
We conclude \sref{s:ahm} with a discussion of interacting fields and local observables in
SQFT in general, and in the Abelian Higgs model in particular.

In \sref{s:comp}, we contrast the $L$-$Q$ and $L$-$V$ variants of the string-localized
approach on the physical Hilbert space with various
alternative approaches, beginning with the standard spontaneous
symmetry breaking (\sref{s:SSB}). They also include the BRST approach
(\sref{s:BRST}) and a ghost-free point-localized approach in Krein
space (\sref{s:Krein}). In both of them, the
restoration of Hilbert space positivity is the principle that
fixes higher orders of the interaction. The latter, however, turns out to be inconsistent
in third order. We also present another string-localized
ghost-free approach on the Krein space. All these alternative approaches need unphysical
  field degrees of freedom.

While the main result of \sref{s:ahm} was the equivalence between a
renormalizable string-dependent interaction and a non-renormalizable
point-localized interaction (which gives rise to a prescription to
renormalize the latter), we present in \sref{s:syn} strong evidence that the
latter is also equivalent to the (equally non-renormalizable)
interaction in the unitary gauge of the gauge-theoretic approach. See
more on the comparison of approaches in \sref{s:syn}.  

A crucial message is that the Higgs potential \eref{VH} is {\em
  the same in all consistent approaches} -- whether they assume a spontaneously
broken classical gauge symmetry, or whether they impose quantum
principles.
We take this as evidence that the observation of the Higgs {\em particle} and
its couplings should not be misconceived as a proof of the Higgs {\em mechanism} as a physical process.

\section{The $L$-$V$ formalism}
\label{s:sloc}

\subsection{String-localized quantum fields}
\label{s:s-qf}

String-localization is the mildest form of relaxing the localization,
bringing substantial benefits. In \cite{MRS2,MRS3}, we have shown how the
string-localized {\em massless} vector potential of QED
\bea{AFe} A_\mu(x,e) = \int_0^\infty ds\, F_{\mu\nu}(x+se)e^\nu,\eea
where $F_{\mu\nu}$ is the Maxwell tensor and $e\in\RR^4$ is a suitable spacelike  string direction,
can be employed for a new understanding of the singular infrared structure
and the ``photon clouds'' of QED, with the usual gauge redundancy
turned into a rich superselection structure.
Other advantages have been discussed in \cite{MRS1}.

In the present paper, the focus is instead on the improved ultraviolet behaviour, i.e.,  the
renormalization of interactions that are non-renormalizable in
Hilbert space formulations of point-local perturbation theory.

In the Abelian Higgs Model, the
string-localized field is given by the same formula \eref{AFe} with
$G_{\mu\nu}=\pa_\mu B_\nu - \pa_\nu B_\mu$, the field strength of the
{\em massive} Proca field $B_\mu$, in the place of $F_{\mu\nu}$. It is defined on the Wigner Fock
space of the Proca field and it creates the same physical particle states
as the latter. It only differs by an
``operator-valued gauge transformation'':
\bea{ABphi} A_\mu(x,e)= B_\mu(x) + \pa_\mu \phi(x,e),\eea where the
massive ``escort field'' $\phi(x,e)$ is given by
\bea{phiBe} \phi(x,e) =
\int_0^\infty ds\, B_\mu(x+se)e^\mu.
\eea
In particular, if smeared with $c(e)$ of total weight 1, $A_\mu(c)$ is another potential for the field strength:
$$G_{\mu\nu}=\pa_\mu A_\nu(c) - \pa_\nu A_\mu(c).$$

The crucial feature is that, thanks to the
integration along the string, its short-distance dimension is 1, while
that of $B_\mu$ is 2. Thus, while the Proca couplings $B_\mu B^\mu H$
and $B_\mu j^\mu$ to the scalar
Higgs field $H$ and to the Dirac current are non-renormalizable, the couplings
$A_\mu(c)B^\mu H$ and $A_\mu(c)j^\mu$ are renormalizable. See \cite{G1} why ``power
counting'' is the appropriate criterium for renormalizability also with
string-localized fields like \eref{AFe}.

\subsection{$L$-$V$ pairs and $L$-$Q$ pairs}
\label{s:LVLQ}

An ``{\bf $L$-$V$ pair}'' is a relation 
  $$L=L'+\pa_\mu V^\mu $$
  between two interaction densities with specific properties, depending on the
  case.
Adding a total derivative to the interaction may be
beneficial. 

E.g., $L$ may
  be defined on a Hilbert subspace of a Krein space where $L'$ is
  defined. This happens in BRST, where the variation of the interaction
  density $L'$ is a total derivative. To make it zero by adding a
  total derivative to $L'$, one may use string-localized free fields. E.g.,
  the escort field of QED on the Krein space satisfies $s(\phi(c)) = u$,
  so that $A(c)=A^K+\pa\phi(c)$ and $L(c)=A(c)j$ are BRST
  invariant. For another example, see \sref{s:LV}.

Or $L$ may be power-counting renormalizable while $L'$ is
  not. This is the situation in the present paper, where both $L$ and $L'$ 
  live on the physical Hilbert
  space of the Abelian Higgs model. Renormalizability of $L$ is achieved by string-localization.

In classical field theory, total derivatives in the Lagrangian are ineffective for the
equations of motion because the total action is the same. In
contrast, in 
quantum field theory, the S-matrix $$S=Te^{i\int d^4x\,L(x)}$$ is the time-ordered exponential of
the action. Because the time-ordering does not commute with
derivatives, adding a derivative 
will in general change the S-matrix.

 The general formalism to be developed below allows to add derivative terms {\em without altering} the
    S-matrix.  It rather {\em provides an equivalent reformulation of a theory}
in which complementary principles are manifestly satisfied, see \eref{SILV}. The
equivalence then shows that all principles hold simultaneously. E.g.,
while a non-renormalizable 
interaction has no autonomous interpretation, it is provided by a renormalizable
(string-localized) reformulation, see \rref{r:rnr}.


The general idea is quite flexible, and can also be used outside SQFT,
whenever derivative terms play a role. In the main body of the paper,
both $L$ and $L'$ live on the physical Hilbert space of the Abelian
Higgs model. This is not possible for QED because of
the vanishing photon mass and the IR problem: instead,
an $L$-$V$ pair reformulating the standard indefinite
(Feynman gauge) Krein space interaction as an (embedded) string-localized Hilbert space
interaction has proven to be a powerful tool for the understanding
of infrared features \cite{MRS3}. In \sref{s:Krein} of this paper,
we consider an $L$-$V$ pair for the Abelian
Higgs model between two point-localized formulations,
one on the Hilbert space and the other on the Krein space.

In \sref{s:PSI-imp}, we develop the $L$-$V$ formalism specifically for a pair of a
point-localized non-renormalizable  and a string-localized
renormalizable interaction. In the general case, only the specific
properties of the admitted interaction terms have to be changed.

When one adds a string-localized derivative term in
order to render a point-localized interaction renormalizable, one has to {\em impose} the {\bf Principle of String Independence} that the S-matrix in higher orders  (and the local observables
of the theory) are independent of the auxiliary string direction.

The implementation of this principle may require just standard Ward identities, as
in QED, or it may require, as in the Abelian Higgs Model, the
addition of further ``induced'' terms to the interaction density in order 
to compensate obstructions against string-independence that occur because time-ordering does not respect differential
relations among the fields. Moreover, cancellation of obstructions in higher
order may fix values of parameters that were free parameters in lower
orders. This is how the Lie-algebra structure of self-interactions of
vector bosons (Yang-Mills), the necessity of a Higgs coupling in the
massive case, and the chirality of the coupling of
massive vector bosons to Fermi fields (weak interaction) were shown
\cite{GGM,GMV}. This is also how the correct coefficients in \eref{VH}
will be fixed in \sref{s:ahm} of the present paper.

String-localized interactions always depart from an 
$L$-$V$ pair, or, slightly more flexible, an $L$-$Q$ pair. The former
consists of a renormalizable string-localized interaction density
$L_1(c)$ and a Lorentz vector of Wick polynomials $V_1^\mu(c)$ such that 
\bea{LVpair}L_1(c)  = \Lp_1+\pa_\mu V_1^\mu(c), \eea
where $\Lp_1$ is a (typically non-renormalizable) point-localized interaction density. 
Here $c$ stands for the dependence on a string direction $e$
that may be smeared with a smearing function $c(e)$. 
The simplest $L$-$V$ pairs are of the form
\bea{LAj} A_\mu(c) j^\mu = B_\mu j^\mu + \pa_\mu(\phi(c)
j^\mu),\eea 
where $j$ is a conserved current, $B_\mu$ is a
point-localized vector potential, and $A_\mu(c) = B_\mu +
\pa_\mu\phi(c)$ is an associated string-localized vector potential of
the form \eref{ABphi} (smeared with $c(e)$). In QED, $B_\mu$ is the
Feynman gauge vector potential defined on a Krein Fock space.

A more flexible version is the $L$-$Q$ pair formalism. It starts from the weaker condition
\bea{LQpair} \delta_c L_1(c) = \pa_\mu Q_1^\mu(c)\eea
which only states that the {\em variation} of the string-localized interaction
density w.r.t.\ the string direction or its smearing function is a
total derivative.
(When $L_1(c)$ belongs to an $L$-$V$ pair, then \eref{LQpair} is trivially
fulfilled with $Q_1^\mu =\delta_c V_1^\mu$.)

Its main advantage is that it does not assume the existence of $V_1^\mu(c)$ such
that $Q_1^\mu(c) = \delta_c V_1^\mu(c)$, nor of an associated point-localized
density $\Lp_1$ such that \eref{LVpair} holds. It therefore does not
reformulate a given non-renormalizable interaction, but
rather allows to establish string-independence of a renormalizable theory whose
interaction is string-localized from the outset.
In the Abelian Higgs Model, already the $L$-$Q$-pair approach allows to fix the
Higgs potential. But strictly speaking only the $L$-$V$-pair approach allows to identify this potential with the Higgs potential of other
approaches, because it contains 
the point-localized interaction to compare with, and thus
is closer to conventional model building. 

Some interactions can be formulated as an
$L$-$Q$ pair on the Wigner Hilbert space of free fields, but not as an
$L$-$V$ pair. An example is massless Yang-Mills on the Hilbert space
of the free field strengths
$F^a_{\mu\nu}$  \cite{G2,GGM}.
Thus, one will have to resort to the $L$-$Q$ formalism for
QCD, if one wants to preserve positivity. See also \sref{s:disc}.

In a way, the $L$-$Q$ equation \eref{SILQ} describes
  only the infinitesimal departure (in first order of $\partial\chi$) from the adiabatic limit
  of the $L$-$V$ identity \eref{SILV}. This results in drastic computational
simplifications. For this reason  we begin the next subsection,
after some preparations,  with the exposition of the former.

\subsection{Implementing the Principle of String Independence}
\label{s:PSI-imp}

\paragraph{Notations.}
In the sequel, we consider fields $X$ as Wick polynomials in a
basis of free fields labeled by $\varphi$. 
In order not to overburden the notation, we shall frequently write
$X^{(\prime)}$ for fields $X(x^{(\prime)})$. Wick ordering is always
understood but not written, except in some lemmas and their proofs where otherwise
there may be ambiguities. The time-ordering symbol
$T$ is meant to apply to all fields to its right (brackets
omitted). $\erw{\,\cdot\,}$ is the free vacuum expectation value. $\delta_{xx'}$ stands for $\delta(x-x')$ and $\delta_{xx'x''}\equiv
\delta_{xx'}\delta_{x'x''}$ for the total $\delta$-function; and
$\SS_n$ for $\frac 1{n!}$ times the sum over all permutations of
$n$ points $x,x',\dots$.

\paragraph{Obstructions.}
Because we are interested in {\em necessary} conditions for
string-independence (see \rref{r:rnr}), we shall
investigate the SI condition only at tree level.

String-independence possibly fails because time-ordering
does not commute with derivatives. It turns out that the PSI 
at tree level can be formulated in terms of ``obstructions'' of the form 
\bea{OVdef}O_Y(X'):=[T,\pa_\mu]Y^\mu X'\tree \equiv T\pa_\mu Y^\mu(x)X(x')\tree -
\pa_\mu TY^\mu(x)X(x')\tree ,\eea
where $Y^\mu$ ($=V^\mu$ or $Q^\mu$, respectively) are vector-valued
fields. The quantities $O_Y(X')$ can be
expanded in terms of the numerical ``two-point
obstructions'' among the basis fields 
\bea{2pt-obs}O_\mu(\varphi;\varphi')\equiv [T,\pa_\mu]
\varphi(x)\varphi'(x')\equiv\erw{T\pa_\mu\varphi(x)\varphi'(x')}-\pa_\mu\erw{T\varphi(x)\varphi'(x')}.
\eea
The latter are $\delta$-functions, or derivatives or string-integrals of
$\delta$-functions to be determined in each model, see \aref{a:prop}. For the Abelian
Higgs Model, they are displayed in  \eref{OH}--\eref{OA}. 
\blemma{l:OVexp} It holds
\bea{OVexp}
O_Y(X') = \sum_{\varphi,\varphi'}O_\mu(\varphi;\varphi') \cdot \wick{\frac{\pa Y^\mu}{\pa\varphi} \,
\frac{\pa X'}{\pa\varphi'}}.\eea
\elemma
\bcor{c:Lb} The maps $X'\mapsto O_Y(X')$ are derivations on Wick polynomials, i.e., one has the Leibniz rule
$$O_Y(\wick{XY}\!\!')= \wick{O_Y(X')Y'} + \wick{X'O_Y(Y')} .$$
\ecor

\noindent {\em Proof of the Lemma:}
By the Wick expansion, because there is only one contraction at tree
level, it holds
$$TY^\mu X'\tree =
\sum_{\varphi,\varphi'}\erw{T\varphi(x)\varphi(x')}\wick{\frac{\pa Y^\mu}{\pa\varphi}(x)\frac{\pa X}{\pa\varphi'}(x')}.$$
Apply the same expansion to $T(\pa_\mu Y^\mu) X'\tree$, where $\pa_\mu
Y^\mu =
\sum_{\varphi}\wick{\frac{\pa Y^\mu}{\pa\varphi}\pa_\mu\varphi}$,
and subtract the expansion of $\pa_\mu TY^\mu X'\tree$. The
result \eref{OVexp} is obtained, because
all terms cancel in which the derivative hits the uncontracted factors
$\wick{\frac{\pa Y^\mu}{\pa\varphi}(x)\frac{\pa X}{\pa\varphi'}(x')}$. \qed

\noindent {\em Proof of the Corollary:}
Obvious, because the maps $X'\mapsto \frac{\pa X'}{\pa \varphi'}$ are
derivations.
\qed

Thus, once the two-point obstructions have been determined in a model,
the computation of $O_Y(X')$ is straightforward. We have automatized it in many
higher-order cases involving iterations of maps $O_Y$, as in
\eref{O3O} or \eref{O3simp}.

\paragraph{$L$-$Q$-pair formalism.}
The $L$-$Q$-pair approach implements the string independence of a
model with a renormalizable string-localized interaction density in
the adiabatic limit when the spacetime cutoff function $\chi$  for the coupling
constant $g$ goes to 1.

We want to find conditions on a string-dependent renormalizable
interaction such that the variation of the tree-level S-matrix with
respect to the string smearing function $c$ vanishes in the adiabatic limit:
\bea{SILQ} \lim_{\chi\to1}\delta_c T e^{iL[\chi,c]}\tree
\stackrel!=0.\eea
Here $L[\chi,c]$ is a series of the form 
\bea{L}L[\chi,c] \equiv \int dx\, \Big(g\,\chi(x)L_1(x,c) +
\frac{g^2}2\chi^2(x)L_2(x,c) + \dots\Big)\eea
with a sequence $L_1(x,c),L_2(x,c),\dots$ of properly adjusted {\em power-counting
  renormalizable string-localized} interaction densities.

In first order (no time-ordering needed), the condition \eref{SILQ} amounts to the statement
that $\int dx \, \delta_c L_1(x,c)=0$. Thus, $\delta_cL_1(x,c)$ must be
a total derivative of the form \eref{LQpair}, which is therefore always the starting point
of the recursion.

Imposing \eref{SILQ} at tree level order by order, results in a recursive scheme
determining the higher-order interactions $L_n$: in order $n$, the sum
of contributions
from all $L_m$ with $m<n$ may not vanish, and this ``obstruction''
has to be cancelled by $L_n$. Whether this is possible, depends on the model, i.e., on the
``initial data'' $L_1(c),Q_1(c)$. We refer to the
fulfillability of \eref{SILQ} in each order as the condition of string-independence
({\bf SI condition}). The SI condition may induce higher-order
interactions, and at the same time also fix parameters from lower
orders.

After expanding \eref{SILQ} to second order:
$$\frac{i^2}2\int dx\,dx'\, \chi(x)\chi(x')\big(\delta_c TL_1(c)L_1'(c) -i \delta_{x,x'}L_2(c)\big)\tree\stackrel!=0,$$
one may insert 
$\delta_cL_1=\pa Q_1$, and replace $T\pa Q_1L_1'$ by $[T,\pa] Q_1L_1'$
because the subtracted term $\pa TQ_1L_1'$ vanishes in the adiabatic
limit. The resulting
second-order obstruction $O^{(2)}_{LQ}$ must be cancelled by
$\delta_cL_2$ up to another total derivative. This is the second order
SI condition: 
\bea{O2OLQ}O^{(2)}_{LQ}(x,x'):=[T,\pa]Q_1L_1'+[T,\pa']Q'_1L_1=
O_{Q_1}(L_1')+O_{Q_1'}(L_1) \stackrel != i\delta_{xx'}\cdot (\delta_cL_2(x)
- \pa Q_2(x)).\qquad\eea
The condition \eref{O2OLQ} determines $L_2$ and $Q_2$ (possibly with some free
parameters). If $L_1$ is cubic in the fields, then $O^{(2)}_{LQ}$
and the second-order densities $L_2$, $Q_2$ are quartic.

\newpage

After expanding \eref{SILQ} to third order, one may insert 
$\delta_cL_1=\pa Q_1$ and (using \eref{O2OLQ}) $\delta_cL_2=\pa
Q_2 -i\int dx''\, O^{(2)}_{LQ}(x,x'')$. One obtains
\bea{XX}\notag\frac{i^3}6\int
dx\,dx'\,dx''\,\chi(x)\chi(x')\chi(x'')\cdot \hspace{80mm}\\ \notag\cdot\Big(3T\pa
Q_1L_1'L_1''-3i\delta_{xx''}\big(T\pa Q_1L_2' + T\pa Q_2 L_1'\big) -
3 TO^{(2)}_{LQ}(x,x'')L_1'-\delta_{xx'x''}L_3\Big)\stackrel!=0.
\eea
One may again replace $T\pa Q_m\dots$ by
$[T,\pa] Q_m \dots$ wherever it occurs. The resulting term $3\SS_3([T,\pa]Q_1L_1'L_1'')$
can be expanded at tree level, using \lref{l:Lb}:
$$3\SS_3\big([T,\pa]Q_1L_1'L_1''\tree\big) = 3\SS_3\big(2TO_{Q_1}(L_1')L_1''\tree\big),$$
and cancels the
term $\SS_3(TO^{(2)}_{LQ}(x,x'')L_1'\tree)$. The terms that are left define the
third-order obstruction $O^{(3)}_{LQ}$ and should be cancelled by
$\delta_cL_3$ up to another total derivative:
\bea{O3OLQ}O^{(3)}_{LQ}(x,x',x'') := -3i\SS_3\Big(\delta_{x'x''}\big(O_{Q_1}(L_2')+O_{Q_2}(L_1')\big)\Big) \stackrel != \delta_{xx'x''}\cdot(\delta_cL_3-\pa Q_3).\eea
The condition \eref{O3OLQ} determines $L_3$ and $Q_3$. If $L_1$ and $Q_1$ are cubic in free
fields, then $L_n$ and $Q_n$ are of polynomial order $n+2$. The
recursion must stop with $L_3=0$, because
renormalizable $L_n$ of polynomial order $>4$ do not exist.

\paragraph{$L$-$V$-pair formalism.}
The more ambitious $L$-$V$ formalism not only allows to construct {\em
some}
string-independent S-matrix. It also establishes the
equivalence with a possibly non-renormalizable point-localized
interaction {\em before} the
adiabatic limit is taken, see \rref{r:rnr}.

We want to establish the identity
\bea{SILV}Te^{i(L[\chi;c] + \Vc[\chi;c]) }= Te^{i\Lp[\chi]}\eea
to hold at tree level for arbitrary cutoff functions $\chi$, where the term
$\Vcn[\chi]$ vanishes\footnote{The
  notation ``$\Vcn$'' is only suggestive. The main parts of $\Vcn[\chi]$ are
  $V_n^\mu(\pa_\mu(\chi^n))$, see \eref{V}. The need to add such a term in order to get an
  identity {\em before} the adiabatic limit is taken, was first noticed
by Duch \cite{Du}.} in the
adiabatic limit $\chi\to 1$.
More precisely, on the right-hand side
\bea{Lp}
\Lp[\chi] \equiv \int dx\, \Big(g\,\chi(x)\Lp_1(x) +
\frac{g^2}2\chi(x)^2\Lp_2(x) + \dots\Big)\eea
is a series of {\em possibly power-counting
non-renormalizable point-localized} interaction densities. Similarly,
$L[\chi,c]$ on the left-hand side is again given by \eref{L}
with a series of {\em power-counting
renormalizable string-localized} interaction densities. 
Finally,
\bea{V}\Vcn[\chi] =\!\! \int dx\, \Big(g\,\pa_\mu\chi(x)V^\mu_1(x,c) +
\frac{g^2}2\big[\pa_\mu\chi(x)^2V^\mu_2(x,c) +
\pa_\mu\chi(x)\pa_\nu\chi(x)W^{\mu\nu}_2(x,c)\big] + \dots\Big) \qquad
\eea
with a series of string-localized, possibly non-renormalizable
tensor densities $V_n^\mu, \dots, W_n^{\mu_1\dots\mu_n}$.

\brem{r:rnr} The virtue of the formula \eref{SILV} is that the left-hand side is
renormalizable in the adiabatic limit, where the term
$\Vcn[\chi]$ vanishes. It thus serves to ``renormalize the
  non-renormalizable right-hand side'', which is manifestly
  string-independent and point-localized, by 
  fixing infinitely many renormalization constants appearing in loop diagrams
  in terms of finitely many constants on  the left-hand side. For this
prescription to work, \eref{SILV} must be an identity at tree level.
We therefore shall restrict the analysis to tree level.
\erem

Again, we refer to the fulfillabilty of \eref{SILV} as the {\bf
  SI condition}. In each order $O(g^n)$, it is an equality between
operator-valued distributions evaluated on $\chi^{\otimes n}$. It
constitutes a recursive system, that has to be solved for
$L_n(c),\Lp_n, V_n(c), W_n(c)$ with the specifications as given
above.

\newpage

In first order, the SI condition 
simply reads 
\bea{SI1}\int dx\, \chi(x)\big(L_1(x,c) - \pa_\mu V_1^\mu(x,c)-\Lp_1(x)\big)\stackrel!=0.\eea
Its validity for all $\chi$ is equivalent to the $L$-$V$-pair
condition \eref{LVpair}, which is therefore always the starting point
of the recursion.

In $n$-th order, one collects all terms involving
$\Lp_m,L_m,V_m,W_m$ with $m<n$ in \eref{SILV}, and writes them with the
help of integrations by parts as
$$\frac{i^n}{n!} \int dx_1\dots dx_n \, \chi(x_1)\dots \chi(x_n)\,
O^{(n)}(x_1,\dots,x_n).$$
The ``$n$-th
order obstruction'' $O^{(n)}$ must then be cancelled by the linear
contribution from $\Lp_n,L_n,V_n,W_n$. This condition determines the latter,
possibly with free parameters. 

\bpropo{p:O2LV} The tree-level
obstruction in second order is
\bea{O2O} O^{(2)}(x,x')=\SS_2\Big(2O_{V_1}(\Lp_1{}') +
O_{V_1}(\pa'V_1') - \pa' O_{V_1}(V_1') \Big).
\eea
\epropo
For the proof, see \aref{a:SI}. Because $\chi$ is arbitrary, the SI condition requires the cancellation 
\bea{canc2}O^{(2)}(x,x') \stackrel!=i\delta_{xx'}\cdot \big(L_2(x,c)-\Lp_2(x)-\pa_\mu
V_2^\mu(x,c)\big)
+\pa_\mu\pa'_\nu\big[i\delta_{xx'}\cdot W_2^{\mu\nu}(x,c)\big].\eea
\bpropo{p:O3LV} After cancellation of the second-order obstruction, the tree-level
obstruction in third order is
\bea{O3O}
O^{(3)}(x,x',x'')=\SS_3\Big(O_{V_1}\big(O_{V_1'}(\Lp_1{}''+2L_1'')\big)
-2\pa''O_{V_1}\big(O_{V_1'}(V_1'')\big) +\,3O_{O_{V_1}(V_1')}(\Lp_1{}'') - \qquad\\ \notag
-\,3i \delta_{x'x''}\big(O_{V_1}(L_2')-\pa'
O_{V_1}(V_2')+O_{V_2'}(\Lp_1)\big) +3i \pa''_\nu\delta_{x'x''}\cdot O_{W'^\nu_2}(\Lp_1) - 3i\pa'\pa''\big[\delta_{x'x''}O_{V_1}(W_2')\big]\Big)\hspace{-5mm}
\eea
with obvious contractions of Lorentz indices (and
$O_{W_2^\nu}(\Lp_1{}'')\equiv [T,\pa_\mu]W_2^{\mu\nu}\Lp_1{}''$).
\epropo
For the proof, see \aref{a:SI}.
It is interesting to notice, that all terms in
\eref{O3O} are various iterations of  expressions of
the form $O_Y(X')$ as in \eref{OVdef}.
The ensuing SI condition is 
\bea{canc3}O^{(3)}(x,x',x'')\stackrel!= \delta_{xx'x''}\cdot\big(L_3(x)-\Lp_3(x)-\pa_\mu
V_3^\mu(x)\big)
+\SS_3\big(\pa_\mu\pa'_\nu\big[\delta_{xx'x''}\cdot W_3^{\mu\nu}(x)\big]\big).\eea
If the initial $L$-$V$ pair is cubic in free
fields, then the higher-order densities are of polynomial order
$n+2$. Renormalizability requires that $L_3=0$.

\section{The Abelian Higgs Model}
\label{s:ahm}
The basic field content of the Abelian Higgs Model in the SQFT
formulation is given by the fields $A_\mu(c)$ and $\phi(c)=-m^{-2}\pa_\mu
A^\mu(c)$ of mass $m>0$,
and the scalar Higgs field $H$ of mass $m_H>0$. In this spirit, $B_\mu$ (the Proca field) is rather a 
short-hand notation for the string-independent combination
$A_\mu(c)-\pa_\mu\phi(c)$, see \sref{s:s-qf}. Notice that $A_\mu(c)$ and $\phi(c)$ are
defined as in \eref{ABphi} and \eref{phiBe} smeared with $c(e)$ of
total weight 1, so that \eref{ABphi} still holds.

However, for the purpose of the computation of obstructions in the
subsequent analysis, it is more convenient to work in the basis
\bea{basis}  B_\mu , A_\mu(c),\phi(c),H,\pa_\mu
H.\eea
We list $\pa_\mu H$ as
an independent field because time-ordering does not respect differential relations among fields. In contrast, $\pa_\mu\phi(c)=A_\mu(c)-B_\mu$ can be expressed
in terms of the basis fields.

\newpage

We denote the string variation of the
escort field by \bea{wdef} w:=\delta_c\phi(c).\eea Its precise form is
not relevant here; see, e.g., \cite{MRS1}. Then,
\bea{dA} \delta_c\pa_\mu\phi(c)=\delta_cA_\mu(c) = \pa_\mu w, \qquad \delta_cH=\delta_c\pa_\mu H=0.\eea

\subsection{First order}
\label{s:first}
We determine the 
initial $L$-$Q$ and $L$-$V$ pairs that define the model.
We shall see that self-interactions of the massive vector field are
only possible with the intervention of the scalar Higgs field.

Because $A_\mu(c)$, $\phi(c)$ and $H$ have short-distance dimension 1, the
only renormalizable couplings are 
cubic or quartic Wick polynomials, involving at most one derivative in
the cubic case. The most general candidate for
$L_1(c)$ is
\bea{L1cand} L_1 &=& a_1\, A^\mu A^\nu \pa_\mu A_\nu + a_2\, A^\mu
A_\mu\phi + a_3 \, A^\mu \phi\pa_\mu\phi  + a_4\, \phi^3 + \\ \notag
&+&b_1\, A^\mu A_\mu H + b_2\, A^\mu \pa_\mu\phi H + b_3\, A^\mu \phi
\pa_\mu H +b_4\, \phi^2H + \\ \notag
&+& c_1\, A^\mu H \pa_\mu H + c_2\, \phi H^2 + d H^3 +e_1(A^\mu
A_\mu)^2 + e_2\, A^\mu A_\mu \phi^2 + e_3\,\phi^4 +\\ \notag
&+&   f_1\, A^\mu A_\mu \phi H + f_2\, \phi^3 H +g_1\, A^\mu A_\mu H^2
+ g_2\, \phi^2 H^2 + h\, \phi H^3 + j\, H^4.
 \eea
Here, we suppress the string-dependence
of the fields.

 \bpropo{p:LVpair} The interaction density $L_1(c)$ is part of an
$L$-$Q$ pair if, and only if, it is (up to a global factor to be
absorbed in the coupling constant $g$) of the form
\bea{L1}L_1= m\Big(A^\mu B_\mu H + A^\mu\phi \pa_\mu H -
\frac{m_H^2}2\phi^2H+a\,H^3\Big)+a'\,H^4 +\pa_\mu\sum\nolimits_i\alpha_i U_i^\mu,\eea
where $U_1^\mu =
A^\mu A^\nu A_\nu$, $U_2^\mu = A^\mu \phi^2$, $U_3^\mu = A^\mu
\phi H$, $U_4^\mu = A^\mu H^2$. At this point, $a$, $a'$ and
$\alpha_i$ ($i=1,2,3,4$) are free real parameters. 
It holds $\delta_cL_1(c)=\pa_\mu Q_1^\mu(c)$ and 
$Q_1^\mu(c)=\delta_cV_1^\mu(c)$ with
\bea{Q1}
Q_1^\mu&=& m\big(B^\mu wH+\phi w\pa^\mu H\big) + \sum\nolimits_i\alpha_i\delta_cU_i^\mu,\\
\label{V1}
V_1^\mu&=& m\Big(B^\mu\phi H +\frac12\phi^2\pa_\mu H\Big) + \sum\nolimits_i\alpha_iU_i^\mu,
\eea
hence $L_1(c)$
is also part of an $L$-$V$ pair $L_1(c)-\pa_\mu V_1(c) =\Lp_1$ with 
\bea{Lp1}\Lp_1=m\big(B^\mu B_\mu H+ a \, H^3\big)+ a'
\, H^4.\eea
\epropo

\brem{r:alphaab}
We shall show in \lref{l:noalpha} that the SI condition in second order requires
$\alpha_i=0$ ($i=1,2,3,4$.) The term $aH^3$ in \eref{L1},  \eref{Lp1}
will become the cubic part of the potential
\bea{VHab} V(H)= \frac12 m_H^2 H^2- g\cdot ma H^3 - \frac {g^2}2\cdot
bH^4 .\eea
The quartic part will arise in second order in $L_2$ (whereas $a'$ as part of $L_1$
must vanish),
and the coefficients $a$ and $b$ will be fixed in third order, see
below. 
That  the SI condition uniquely fixes the Higgs potential as in
\eref{VH}, is the result referred to in the title of this
paper. 
\erem
\noindent {\em Proof of \pref{p:LVpair}:} We must fix the coefficients in \eref{L1cand} such that the
string-variation $\delta_cL_1(x,c)$ is a total $x$-derivative. The terms
$dH^3+jH^4$ in \eref{L1cand} are string-independent. The four
combinations $\pa_\mu U_i^\mu$ trivially satisfy this condition via
$\delta_c\pa_\mu U_i^\mu = \pa_\mu (\delta_cU_i^\mu)$. We also have
(using \eref{wdef}, \eref{dA}, and 
$\square H=-m_H^2 H$ and $\pa A=\square \phi = -m^2\phi$)
$$\delta_c(A^2H+ A\phi\lrpa H -\frac12m_H^2\phi^2H) = (A\pa H-m_H^2\phi H)w
+ (AH + \phi \lrpa H)\pa w = \pa [(AH-\phi\lrpa H)w].$$

\newpage

Since $A-\pa\phi=B$, these are the solutions displayed in \eref{L1}
and \eref{Q1}, with a
relabelling $d\to ma$, $j\to a'$ of the coefficients. If the coupling
constant $g$ is dimensionless, then $b_1$ is dimensionful. The choice
$b_1=m$ in \eref{L1} is a matter of convenience.

To prove that there are no further solutions, we may use the given
solutions to freely adjust the coefficients $a_1,a_3,b_1,b_3,c_1$. For
an independent solution we
may thus assume $a_1=a_3=b_1=b_3=c_1=0$. By homogeneity in the Higgs
fields and in the vector boson fields, the conditions on the remaining
coefficients decouple from each other  
for the terms with coefficients labelled by different letters. E.g., the
remaining $a$-terms give
$$\delta_c (a_2 A^2\phi + a_4\phi^3) = a_2(A^2w+2(A\pa w)\phi) +
3a_4\phi^2 w,$$
which cannot be a total derivative unless $a_2=a_4=0$. The remaining 
$b$-terms give
$$\delta_c (b_2A\pa\phi H+b_4\phi^2H) = b_2(A+\pa\phi)\pa wH + 2b_4\phi wH,$$
which cannot be a total derivative unless $b_2=b_4=0$. Similar for the terms with
coefficients $e_i$, $f_i$, $g_i$, $h$, which do not admit combinations whose
string-derivatives are a total $x$-derivative. 

The asserted equalities $\delta_c V_1=Q_1$ and $L_1-\pa V_1=\Lp_1$ are verified by direct computation. \qed

\subsection{Two-point obstructions}
\label{s:obs}
In order to compute the higher obstructions $O^{(n)}_{LQ}$ of the S-matrix, one needs the
two-point obstructions involving the fields \eref{basis} (and $w$ in the $L$-$Q$-pair approach) of the 
Abelian Higgs Model. By \eref{OVdef}, the latter are directly
obtained from the propagators as determined in
\aref{a:prop}, in combination with the field equations. The given scaling
degrees of the propagators allow two free renormalization parameters
$c_H$, $c_B$ in the Higgs and Proca sector, respectively. The
relevant two-point obstructions for the Higgs field are
\bea{OH}
O_\mu(\pa^\mu H;\pa'_\nu H')&=& -i(1+c_H) \pa_\nu \delta(x-x'),
\\ \notag
O_\mu(\pa^\mu H;H') &= &i \delta(x-x'),
\\ \notag
O_\mu(H; \pa'_\nu H') &=& ic_H \eta_{\mu\nu} \delta(x-x'),
\\ \notag
O_\mu(H;H')=0.
\eea
Those for the fields $B$, $A$, $\phi$ are
\bea{OB}
O_\mu(B^\mu; B_\nu') &=& -i(1+c_B)\cdot m^{-2}\pa_\nu \delta(x-x'),
\\ \notag
O_\mu(B^\mu; \phi') &=&-im^{-2}\delta(x-x'),
\\ \notag
O_\mu(\phi; B_\nu') &=& -ic_B\cdot m^{-2}\eta_{\mu\nu} \delta(x-x'),
\\ \notag
O_\mu(\phi; \phi') &=& 0,
\eea
as well as 
\bea{OA} 
O_\mu(A^\mu;A'_\nu) &=& -i\cdot\big(e_\nu I_e -
(ee')I_eI_{-e'}\pa\big)\delta(x-x'), \\ \notag
O_\mu(A^\mu;\phi') &=& -i\cdot(ee')I_eI_{-e'}\delta(x-x'), \\ \notag
O_\mu(A^\mu;B'_\nu) &=& -i\cdot e_\nu I_e \delta(x-x'),\\ \notag
O_\mu(\phi;A'_\nu)&=&0, \\ \notag
O_\mu(B^\mu;A'_\nu) &=& 0.
\eea
In the $L$-$Q$-approach, one also needs the two-point obstructions
$O_\mu(w;X')$ for $X=A,B,\phi$. These are all found to be zero.

The two-point obstructions $O(A,X')$ are string-localized. They could spoil the SI conditions, because obstructions
of \eref{SILQ} or \eref{SILV} involving string-integrated $\delta$-functions cannot
be cancelled by higher-order densities. Fortunately,
  this does not happen, as we shall see.
\blemma{l:noalpha}
The SI condition (both in the $L$-$Q$ and $L$-$V$-pair approach) requires in second order that in \pref{p:LVpair},
$$\alpha_1=\alpha_2=\alpha_3=\alpha_4=0.$$
\elemma
\noindent {\em Proof:}  $Q_1^\mu$
and $V_1^\mu$ contain the field $A$ only in the terms $\delta_cU_i$ and
$U_i$. By \eref{OVexp}, obstructions with string-integrated
$\delta$-functions can occur in second order only through these terms
in \eref{Q1} and \eref{V1}. Because their coefficients according to
\eref{OVexp} are linearly independent, there can be no
cancellations. These terms must therefore be excluded  altogether: $\alpha_i=0$. \qed

We proceed with the $L$-$Q$-pair approach. The $L$-$V$-pair approach
will be treated in \sref{s:ahmLV}.

\subsection{The $L$-$Q$-pair approach} 
\label{s:ahmLQ}

The initial $L$-$Q$-pair \eref{LQpair} of the Abelian Higgs
Model is specified by \pref{p:LVpair} and \lref{l:noalpha}:
\bea{LQini} L_1 &=& m\Big(ABH + A\phi\pa H - \frac{m_H^2}2\phi^2H + aH^3\Big)
+a'H^4, \\ \notag
Q_1 &=& m\big(BH+\phi\pa H\big)w.
\eea

\paragraph{Second order.}
The string-localized two-point obstructions
$O_\mu(A^\mu;X')$ in
\eref{OA} do not contribute to the second-order obstruction
\eref{O2OLQ} of the S-matrix, because
the field $A$ (or $\pa \phi =A-B$) does not occur in $Q^\mu$. This
feature of the model distinguishes the choice of the ``kinematical''
propagators for the string-localized fields, as discussed in
\aref{a:sloc-2pt}, leaving only the parameters $c_H$ in \eref{OH} and
$c_B$ in \eref{OB} free.

\bpropo{p:SI2LQ} The SI condition in second order \eref{O2OLQ} requires that the
 parameter $a'=0$ in \eref{LQini}. Then the condition is solved by 
\bea{L2LQ} L_2 &=& m^2\Big(\big(3a+\frac{m_H^2}{m^2}\big)\phi^2H^2 -\frac{m_H^2}4
\phi^4 + (1+c_H)\cdot
A^2\phi^2\Big)+ (1+c_B)\cdot A^2H^2 + bH^4 ,\notag \\
Q_2&=&m^2(1+c_H)\cdot A\phi^2w +(1+c_B)\cdot AwH^2.
\eea
The quartic term of the Higgs potential appears with the coefficient
$b$ undetermined.
\epropo
\noindent {\em Proof:} The task is to compute the obstruction $O^{(2)}_{LQ}$ in \eref{O2OLQ}. 
The straightforward computation using \eref{OVexp} with \eref{OH},
\eref{OB} yields
\bea{OQL1}\notag O_{Q_1}(L_1')
\!\!&=&\!\! m^2\Big(\phi w \!\cdot\!\delta_{xx'}\!\cdot\!
\big(AB-\frac{m_H^2}2\phi^2+3aH^2+\frac{4a'}mH^3\big)+ c_H
Bw\!\cdot\!\delta_{xx'}\!\cdot\! A\phi - (1+c_H) \phi w\!\cdot\!\pa\delta_{xx'}\!\cdot\! A'\phi'\Big) \\
\!\!&&\!\! -wH\!\cdot\!\delta_{xx'}\!\cdot\! (A\pa H -m_H^2 \phi H)  -c_B w\pa
H\!\cdot\!\delta_{xx'}\!\cdot\! AH- (1+c_B)wH\!\cdot\!\pa\delta_{xx'}\!\cdot\! A'H'.
\notag
\eea
Adding $O_{Q_1'}(L_1)$ and using \eref{XdY}, one obtains 
\bea{O2LQ}O^{(2)}_{LQ}\!\!&=&\!\! i\delta_{xx'}\cdot m^2\Big[2\big(3a+\frac{m_H^2}{m^2}\big)\phi wH^2 -
m_H^2\phi^3w +(1+c_H)\big(2AB\phi w +(A\lrpa
w)\phi^2\big)\Big] +\notag \\ &&\!\!+i\delta_{xx'}\cdot (1+c_B)\big(-2AwH\pa H +(A\lrpa
w)H^2\big) + i\delta_{xx'}\cdot  8ma'\phi wH^3  =\notag \\
\notag \!\!&=&\!\! i\delta_{xx'}\cdot m^2\Big[\delta_c\Big(\big(3a+\frac{m_H^2}{m^2}\big)\phi^2H^2 -\frac{m_H^2}4
\phi^4 \Big) +(1+c_H)\Big(\delta_c(A^2\phi^2)-\partial(A\phi^2w)\Big)\Big]+\\ &&\!\!+i\delta_{xx'}\cdot 
(1+c_B)\big(\delta_c(A^2H^2)-\partial (AwH^2) \big) + i\delta_{xx'}\cdot ma'\delta_c(\phi^2H^3).\eea
The  term $\phi^2H^3$ has dimension 5 and is not admissible in $L_2$,
hence we must have $a'=0$. Then, $L_2$ and $Q_2$ in \eref{O2OLQ} are read off
\eref{O2LQ}. \qed

\newpage

\paragraph{Third order.}

$Q_2$ consists of two terms involving the field $A$ with
coefficients $1+c_H$ and $1+c_B$. By \eref{O3LQ} and \eref{OA}, these would
contribute string-integrated $\delta$-functions in $O^{(3)}_{LQ}$, which cannot be
cancelled. This forces us to fix the renormalization parameters as
\bea{ccH} c_H=-1, \qquad c_B =-1.\eea
In particular, $Q_2=0$ with this choice.

\bpropo{p:SI3LQ} The SI condition in third order \eref{O3OLQ} requires that the
 parameters $a$ in \eref{LQini} and $b$ in \eref{L2LQ} take the values
\bea{ab} a=-\frac 12\frac{m_H^2}{m^2}, \qquad b= -\frac14\frac{m_H^2}{m^2}.\eea
Then the condition is solved by $L_3=0$ and $Q_3=0$.
\epropo
\bcor{c:VH} The values $a$ and $b$ determined by \pref{p:SI3LQ} yield the precise form of the
Higgs potential \eref{VH}.
\ecor
\noindent {\em Proof of the proposition:}
The task is to compute $O^{(3)}_{LQ}$ as in \eref{O3OLQ}. 
The straightforward computation yields\footnote{\label{f:indep}We do not know the significance of the following
  observation. If one works with general $c_B$ and $c_H$, then one
  gets -- apart from the string-localized
  obstructions, that do not cancel -- four 
  additional point-localized contributions to \eref{O3LQ}
$$+  2\big((1+c_H)+c_B(1+c_H)-(1+c_B)-c_H(1+c_B)\big)A^2\phi wH =0,$$
which identically cancel each other. This is a remarkable independence
of the renormalization parameters. In particular, the values of the
Higgs potential parameters $a$
and $b$ are not affected.}
\bea{O3LQ} O^{(3)}_{LQ}=3\delta_{xx'x''}\cdot
\Big[m^3\Big(2\big(3a+\frac{m_H^2}{m^2}\big)+\frac{m_H^2}{m^2}\Big)\phi^3wH +
m\Big(4b-2\big(3a+\frac{m_H^2}{m^2}\big)\Big)\phi wH^3
\Big].\eea
Since the fields $\phi^3wH$ and $\phi
wH^3$ cannot be written as $\delta_c L_3 - \pa Q_3$ with
renormalizable $L_3$, their coefficients
must vanish. This fixes the parameters $a$ and $b$, and
  $\delta_c L_3 - \pa Q_3=0$.
 \qed

\noindent {\em Proof of  the corollary:} 
\eref{VHab} with $a$ and $b$ as in \eref{ab} is \eref{VH}.
\qed

\subsection{$L$-$V$-pair approach}
\label{s:ahmLV}

The initial $L$-$V$-pair \eref{LVpair} of the Abelian Higgs
Model, as specified by \pref{p:LVpair} and \lref{l:noalpha}, is 
\bea{LVini} \Lp_1&=&m\big(B^\mu B_\mu H+ a \, H^3\big)+ a'
\, H^4, \\ \notag L_1 &=& m\Big(A^\mu B_\mu H + A^\mu\phi\pa_\mu H - \frac{m_H^2}2\phi^2H + aH^3\Big)
+a'H^4, \\ \notag V_1^\mu &=& m\Big(B^\mu\phi H +\frac12\phi^2\pa^\mu H\Big) .\eea

\paragraph{Second order.}
The string-localized two-point obstructions
$O_\mu(A^\mu;X')$ in
\eref{OA} do not contribute to the second-order obstruction
\eref{O2O} of the S-matrix, because
the field $A$ does not occur
in $V^\mu$. 

 \bpropo{p:SI2} The SI condition in second order \eref{canc2} requires that the
 parameter $a'=0$ in \eref{LVini}. Then the condition is solved by 
\bea{L2} \Lp_2 \!\!&=&\!\! -3B^2H^2 + (1+c_B)\cdot 4B^2H^2 +bH^4, \\ \notag
L_2\!\!&=&\!\! m^2\Big((3a+\frac{m_H^2}{m^2})\phi^2H^2-\frac{m_H^2}4\phi^4
+(1+c_H)\cdot A^2\phi^2\Big)+
(1+c_B)\cdot A^2H^2 + bH^4 , \\ \notag
V_2^\mu \!\!&=&\!\! -B^\mu\phi H^2 +\frac{m^2}6B^\mu\phi^3  + (1+c_H)\cdot \frac
{m^2}2A^\mu\phi^3 + (1+c_B)\cdot A^\mu\phi H^2 ,
\\ \notag
W_2^{\mu\nu}\!\!&=&\!\! \Big((1+c_H)\cdot \frac {m^2}4\phi^4 +
(1+c_B)\cdot \phi^2H^2\Big)\eta^{\mu\nu} .
\eea
with one new free parameter $b$.  
\epropo
{\em Proof:}
We have to compute \eref{O2O}, using \eref{OVdef}. We begin with the
choice \eref{ccH} for $c_H$,
 $c_B$. With this choice, the obstructions $O_V(X')$ contain no
 derivatives or string-integrals of
 $\delta(x-x')$, and it is convenient
to write
\bea{Omdef} O_V(X')=:i\delta(x-x')\cdot \Omega_V(X).\eea
In particular, one has $\Omega_{V_1}(V_1)=0$. This simplifies
\eref{O2O} and \eref{canc2} to
\bea{O2simp}O^{(2)}(x,x') = i\delta(x-x')\cdot
\Omega_{V_1}(\Lp_1+L_1), \qquad
\Omega_{V_1}(\Lp_1+L_1)\stackrel!=L_2-\Lp_2-\pa V_2. \eea
The straightforward computation yields
\bea{OVLpL}\notag \Omega_{V_1}(\Lp_1+L_1) &=&AB H^2 +2B^2H^2+2B\phi
  H\pa H + \\ \notag &&+ m^2\Big(\big(3a+\frac{m_H^2}{m^2}\big)\phi^2H^2
 -\frac12 (A-B)B\phi^2-\frac{m_H^2}4\phi^4\Big) +4ma'\phi^2H^3, \eea
which can be re-written as
$$=3B^2H^2 +m^2\Big(\big(3a+\frac{m_H^2}{m^2}\big)\phi^2H^2 -\frac{m_H^2}4\phi^4\Big)+ 
\pa\big(B\phi H^2
-\frac{m^2}6B\phi^3\big)  +4ma'\phi^2H^3.$$
The first term is point-localized, the second term is
renormalizable, and the third term is a derivative. The last term is not compatible with
the required form of the cancelling second-order densities in \eref{O2simp}. Thus $a'=0$.  
Comparing the other terms with \eref{canc2}, one reads off \eref{L2}
for the special values
$c_H=c_B =-1$.

The additional contributions to $O^{(2)}(x,x')$ due to
different values of $c_H$, $c_B$ involve derivatives of
$\delta(x-x')$. The formulae in
\lref{l:2delta} in \aref{a:useful}
nicely deal with the symmetrization in $x\lra x'$, and reduce the
result to the additional terms
\bea{deltaO2}
\dots  +
(1+c_H) \cdot  im^2\Big(\delta(x-x')\big(A^2\phi^2 -\frac12\pa(A\phi^3)\big)+ 
\frac 14\pa\pa'\big[\delta(x-x')\phi^4\big]\Big) +\qquad\quad \notag \\ \notag +\,(1+c_B)\cdot i\Big( \delta(x-x')\big(A^2H^2-4B^2H^2 -\pa
(A\phi H^2)\big) +
\pa\pa'\big[\delta(x-x')\phi^2H^2\big]\Big).
\eea
This yields the additional terms 
displayed in \eref{L2}. \qed

$L_2$ in \eref{L2} coincides with $L_2$ in \eref{L2LQ} in the
$L$-$Q$-pair approach. Notice that, while $\delta_c V_1=Q_1$, the general setup does not imply that
$\delta_c V_2$  should equal $Q_1$.

\paragraph{Third order.}

$V_2$ contains two terms involving the field $A$ with
coefficients $1+c_H$ and $1+c_B$. By \eref{O3O} and \eref{OA}, these would
contribute string-integrated $\delta$-functions in $O^{(3)}$, which cannot be
cancelled. This forces us again to fix the renormalization parameters
as in \eref{ccH}. In
this case,  the two-point obstructions \eref{OH} and \eref{OB} do not contain derivatives of
$\delta$-functions and can be cancelled in each order by induced
triples $L_n,V_n,\Lp_n$ (i.e., all $W_n=0$). In
particular, one has \eref{Omdef} with
$\Omega_{V_1}(V_1)=0$, so that  $W_2=0$ and \eref{O3O} simplifies considerably:
\bea{O3simp}O^{(3)}(x,x',x'') =
\delta_{xx'x''}\big(-\Omega_{V_1}^2(\Lp_1+2L_1)
+3\Omega_{V_1}(L_2) +3\Omega_{V_2}(\Lp_1) -2\pa \Omega_{V_1}(V_2)\big).\quad\eea

\bpropo{p:SI3} The SI condition in third order \eref{canc3} requires that the
 parameters $a$ in \eref{LVini} and $b$ in \eref{L2} take the values
\eref{ab}. 
In this case, it is solved by
\bea{L3}
\Lp_3&=& \frac{12}m B^2H^3, \\ \notag L_3&=&0, \\ \notag
V_3 &=& \frac2m B\phi H^3-\frac1m\phi^2H^2\pa
  H -\frac{5m}3B\phi^3H+\frac m{12}\phi^4\pa
  H, \\ \notag W_3 &=& 0. \eea
 \epropo
{\em Proof:}
The computation of \eref{O3simp} is straightforward. The result,
inserted in the third-order SI condition \eref{canc3}, can be written as
\bea{O3final} L_3-\pa V_3-\Lp_3 &\stackrel!=&
-\frac{12}m B^2H^3-\pa\Big(\frac2m B\phi H^3-\frac1m\phi^2H^2\pa
  H -\frac{5m}3B\phi^3H+\frac m{12}\phi^4\pa H\Big)
 +\notag \\ &&+m\Big(6b-9a-3\frac{m_H^2}{m^2}\Big)\phi^2H^3
+m^3\Big(\frac{9a}2+\frac94\frac{m_H^2}{m^2}\Big)\phi^4H,\eea
where $L_3$ must vanish because renormalizable fifth-order Wick
polynomials do not exist. The last two terms in \eref{O3final} are not compatible with
the required form of the left-hand side. Their vanishing requires the
values given in \eref{ab}, and hence implies the precise form of the
Higgs potential \eref{VH}, as in \cref{c:VH}.

The first term is a total derivative and should be identified with
$-\pa V_3$. The second term is string-independent, and should be
identified with $-\Lp_3$. These are the third-order terms \eref{L3}. \qed

The renormalizable string-localized interaction density $L[\chi;c]$ in
\eref{L} terminates with the quartic
terms $L_2$, because renormalizable candidates of higher polynomial
order do not exist. Higher-order terms appear only in the form of 
$\Lp_n$ and $V_n(c)$. Recall that the purpose of the (non-renormalizable)
derivative terms $V_n^\mu(\pa_\mu\chi^n)$ is to dispose of the 
non-renormalizable contributions of the point-localized interaction
$\Lp_n(\chi^n)$ in \eref{SILV}, and that they vanish in the adiabatic limit.

\subsection{Coupling to Dirac fields}
\label{s:fermi}

The Abelian Higgs Model serves as a simplified model for the
self-coupling of massive vector bosons in the weak
interaction, when a cubic self-interaction of a single massive vector
field is not viable, see the discussion in \sref{s:intro} and
\pref{p:LVpair}. The model can easily be extended to include also the coupling to
a fermionic current $j^\mu=\ol\psi\gamma^\mu\psi$. Namely, the
interaction \eref{LAj} is another $L$-$V$ pair that can be added to the
$L$-$V$ pair \eref{LVini} of the Abelian Higgs Model.

In order to compute the effect of the extension on the SI condition,
one proceeds as before, using that the obstruction $O_\mu(j^\mu;j'^\nu) =
-\pa_\mu Tj^\mu(x) j^\nu(x')$ vanishes by the usual Ward
identity. With
$$\Delta L_1= Aj,\qquad \Delta \Lp_1= Bj,\qquad \Delta
V_1= \phi j,$$
one finds that the second and third order SI conditions are satisfied with
$$\Delta L_2= 0,\qquad \Delta \Lp_2= -4m\inv \cdot 
BHj-m^{-2}\cdot j^2,\qquad \Delta V_2=
-m\inv \cdot \phi H j,$$
$$\Delta L_3= 0,\qquad \Delta \Lp_3= 18 m^{-2} \cdot BH^2j + 6m^{-3}
\cdot Hj^2,\qquad \Delta V_3= 2m^{-2} \cdot \phi
H^2j - \frac23\cdot \phi^3j.$$
In particular, the Higgs potential is not affected by the
extension. This was expected, because the parameters $a$
and $b$ are fixed at tree level, whereas diagrams 
involving Dirac fields, that could possibly contribute to the
coefficients of  $\phi^3H^2$ and $\phi H^4$, must
necessarily contain Dirac loops.

\subsection{Local observables}
\label{s:locobs}
The ``off-shell'' interacting quantum
  fields are of prime interest for the perturbative construction of an actual
  QFT, beyond the S-matrix needed for predictions of experiments. Not
  all of them are local observables; e.g., in BRST the observables are by
  definition those fields that commute with the interacting BRST
  operator. They are, however, usually not computed in the literature
  \cite{ADS,Scha}.

Interacting fields
are computed in causal perturbation theory by the
variation of ``relative S-matrices'' with respect to a source function
\bea{Bog}\Phi\big\vert_{L(\chi)} (x):= -i \frac{\delta }{\delta f(x)}
S(\chi,0)^* S(\chi,f)\big\vert_{f=0}\eea
where $S(\chi,f) = T e^{i (L(\chi)+ \Phi(f))}$. By axiomatizing
properties of relative S-matrices, this is the way to give a precise
meaning (as the adiabatic limit $\chi\to1$ of \eref{Bog}) to Bogoliubov's formula
$$\Phi\big\vert_{L} (x):= \big(Te^{i\int dx\, L(x)}\big)^*\, 
T \big(\Phi(x) e^{i\int dx\, L(x)}\big).$$
In the $L$-$V$ approach at hand, the formula has to be qualified: 
We shall first establish the identity
\bea{SIXLV}
T\big(\Phi(x)e^{i(L[\chi;c] + \Vc[\chi;c]) }\big) = T\big(\Phi_{[g\chi]}(x)e^{i\Lp[\chi]}\big).
\eea
\eref{SIXLV} is \eref{SILV} with the insertion of a local free field $\Phi(x)$
on the left-hand side, and of
\bea{Xchi}\Phi_{[g\chi]}(x) = \Phi(x)+ g\chi(x)\Phi_{[1]}(x) + \frac
{g^2}2\chi(x)^2\Phi_{[2]}(x)+ \dots\eea
on the right-hand side. The corrections $\Phi_{[n]}(x)$ are free Wick
polynomials that are
recursively determined by \eref{SIXLV}, see below.

The combination of \eref{SILV} and \eref{SIXLV} yields
\bea{XLV} \big(Te^{i(L[\chi;c] +
  \Vc[\chi;c]) }\big)^*\,T\big(\Phi(x)e^{i(L[\chi;c] + \Vc[\chi;c]) }\big)
= \big(Te^{i\Lp[\chi]}\big)^*\,T\big(\Phi_{[g\chi]}(x)e^{i\Lp[\chi]}\big).\eea
Either of these two expressions (to be taken in the adiabatic limit
$\chi\to1$) defines the interacting field
$\Phi\big\vert_L$, where the left-hand side is renormalizable, and the
right-hand side is local {\em if and only if $\Phi_{[g\chi]}(x)$ is
  point-localized}. As in \rref{r:rnr}, infinitely many
renormalization constants on the right-hand side are fixed as functions of finitely many constants of  the left-hand side.

  \bdefi{obs} A free Wick polynomial $\Phi$ such that $\Phi_{[g\chi]}$ is
  point-localized (i.e., $\Phi$ and all its corrections $\Phi_{[n]}$
  are point-localized), is called the
  ``seed'' of the local interacting field $\Phi\vert_{L}$ given by \eref{XLV}. The
  resulting interacting fields are the local observables of the theory.
\edefi

In other words: {\em The perturbation theory selects the local observables of the
  theory}. The condition (vanishing of $\Phi_{[n]}$) can be decided at
the level of the free field.

\newpage

\eref{XLV} generalizes the construction of the interacting Dirac 
field of QED as a point-localized perturbation of the string-localized
``dressed Dirac field'' \cite[Eq.\ (2.14)]{MRS3}. In QED, the SI
condition is fulfilled without any higher order interactions 
added to the massless $L$-$V$ pair \eref{LAj}. In the case at hand,
the role of the dressing transformation is taken by the map
$\Phi\mapsto \Phi_{[g\chi]}$ when $\chi\to1$.

We now turn to the determination of the correction terms
$\Phi_{[n]}(x)$ in \eref{Xchi}.
The strategy is the same as for the S-matrix, cf.\ \aref{a:SI}. We sketch it here again in
a model-independent way, but with the simplifying assumption that the
two-point obstructions involve no derivatives of
$\delta$-functions (otherwise, one would
have to admit terms with derivatives of $\chi$ in \eref{Xchi}), and $\Omega_{V_1}(V_1)=0$. Recall
that in the Abelian Higgs model, the SI condition forces us to choose
the renormalizations \eref{ccH} such that
the simplifying assumptions hold, and as a consequence $W_2=W_3=0$.

Expanding both sides of \eref{SIXLV}
to first order in $g$, we get
$$ i\int dy\, \big(\chi(y)TL_1(y)\Phi(x)+ \pa\chi(y)
TV_1(y)\Phi(x)\big) = i\int dy\, \big(\chi(y)T\Lp_1(y)\Phi(x) +\chi(x)\Phi_{[1]}(x)\big).$$
Inserting $L_1=\Lp_1+\pa V_1$, we get
\bea{Phi1} i\int dy\, \chi(y)[T,\pa^y]
V_1(y)\Phi(x) =\chi(x)\Phi_{[1]}(x)  \quad \stackrel{\eref{Omdef}}\RA
\quad \Phi_{[1]} = -\Omega_{V_1}(\Phi).\eea
Expanding \eref{SIXLV} in second order, we insert
$L_1=\Lp_1+\pa V_1$. This cancels the terms involving the
cubic time-ordered products
$T\Lp_1(y)\Lp_1(y')\Phi(x)$. The remaining cubic terms are of the form 
$O_Y(X',\Phi(x))$ as in \eref{OVXY} and can be evaluated using \lref{l:Lb}. 
This produces terms $2
T\Lp_1\Omega_{V_1}(\Phi(x))+T\Phi(x)\Omega_{V_1}(2\Lp_1+\pa V_1)$,
which cancel the quadratic contributions $T\Lp_1\Phi_{[1]}$ and $T(L_2-\pa V_2-\Lp_2)\Phi(x)$. The remaining
terms give the surprisingly simple result:
\bea{Phi2} \Phi_{[2]} = \Omega_{V_1}\circ\Omega_{V_1}(\Phi) -
\Omega_{V_2}(\Phi).\eea
If the simplifying assumptions as above are fullfilled, then we
conjecture for higher orders:
\bconj{cj:corr}
All corrections $\Phi_{[n]}$ are linear
combinations of iterated obstructions $\Omega_{V_{n_1}}\circ\dots\circ\Omega_{V_{n_k}}(\Phi)$ with $\sum_k
n_k=n$.
\econj
That $\Omega_{V_n}$ should come in $\Phi_{[n]}$ with the coefficient $-1$, can be seen
rather easily. 
Specifically, for reasons to be explained below and after \eref{strange}, we guess\footnote{The effort
  required for the recursive 
  analysis of \eref{SIXLV} in third order is comparable to \eref{SILV} in fourth order.}  
\bea{Phi3} \Phi_{[3]} = \big(-\Omega_{V_1}\circ\Omega_{V_1}\circ\Omega_{V_1} + 2
\Omega_{V_1}\circ \Omega_{V_2}+  \Omega_{V_2}\circ \Omega_{V_1} -
\Omega_{V_3}\big)(\Phi).\eea
This guess, together with \eref{Phi1} and \eref{Phi2} and the
  derivation property of $\Omega_{V_n}$, implies
$$(XY)_{[2]}= X_{[2]}Y+2X_{[1]}Y_{[1]}+XY_{[2]}, \qquad (XY)_{[3]}=
X_{[3]}Y+3X_{[2]}Y_{[1]}+3X_{[1]}Y_{[2]}+ XY_{[3]}.$$
This structure in turn entails that the point-locality of the corrections of two
fields passes to the corrections of their Wick product, hence it
warrants that the seeds $\Phi$ of local interacting fields in \dref{obs} form an
{\em algebra}.\footnote{This is a desirable feature, but not an axiom
  because the map $\Phi\mapsto\Phi_{\rm int}$ in \dref{obs} must not be expected to be
an algebra homomorphism.} However, this feature only constrains, but does
not fix the coefficients in \eref{Phi3}. 

\newpage

The actual computations in the Abelian Higgs model are again
straightforward,  using \eref{OVexp} and the known two-point
obstructions. The first corrections 
$$B^\mu_{[1]}= -\frac1m(B^\mu H+\phi\pa^\mu H), \qquad H_{[1]}= -\frac m2\phi^2, \qquad (\pa_\mu H)_{[1]}= m B_\mu\phi, $$ of the
Proca and the Higgs fields 
are string-localized.  Thus neither the interacting Higgs field nor
the interacting Proca field are local observables of the model.

For the massive Proca field strength tensor $G_{\mu\nu}=\pa_\mu
B_\nu-\pa_\nu B_\mu$, one needs also two-point obstructions involving
$G_{\mu\nu}$. The kinematical propagators
displayed in \aref{a:sloc-2pt} yield
\bea{OG} O_\kappa(B^\kappa, G_{\mu\nu}') = O_\kappa(\phi, G_{\mu\nu}') = 0,
\qquad O_\kappa(A^\kappa, G_{\mu\nu}') = -i I_e
e_{[\mu}\pa_{\nu]}\delta(x-x').\eea
Thus, the first two (three, if the above structural conjecture is correct) corrections of $G_{\mu\nu}$ are zero for the
simple reason that the field $A$ does not occur in $V_1$ and $V_2$
(and $V_3$).

For a
Wick polynomial $\Phi$ in $B$, $H$, and $\pa H$ to be the seed of a
local interacting field, $\Phi_{[1]}=-\Omega_{V_1}(\Phi)$ must be
independent of $\phi$. This condition is a differential equation for
$\Phi$, which implies that $\Phi$ must be a polynomial in the 
composite field $$Z:=m^2B^2+(\pa H)^2.$$
Indeed, also the next two corrections of $Z$ are point-localized:
\bea{strange}Z_{[1]} = -2mB^2H, \qquad Z_{[2]} = 6B^2H^2,\qquad
Z_{[3]} = -\frac{24}m B^2H^3.\eea
$Z_{[3]}$ was computed with \eref{Phi3}, which is the unique 
(up to a factor) combination with a point-localized outcome. We take
this as a strong support for the
guess \eref{Phi3}. Clearly, a better understanding of the higher-order
corrections is strongly desired.

With these evidences, we conjecture
\bconj{cj:locobs}
The 
interacting fields $G_{\mu\nu}$ and $Z=m^2B^2+(\pa H)^2$ are local
observables of the Higgs model.
\econj
Other fields, like the interacting Higgs field, may still be
  regarded as part of the
theory, e.g., in order to create Higgs particle states. But, just as
the interacting dressed Dirac field of QED \cite{MRS3}, they
cannot be local fields in the sense of the usual QFT axiomatics.

When a Dirac field is added to the Abelian Higgs model (see \sref{s:fermi}), one expects the
current to be a local observable. Indeed, all
corrections $j_{[n]}$ vanish because $O_\mu(j^\mu;j'^\nu)=0$.  
The corrections to the Dirac field are computed with the help of
$$O_\mu(j_\mu;\psi') = -\pa_\mu Tj^\mu(x) j^\nu(x') =\delta_{xx'}\psi(x),$$ giving
$$\psi_{[1]}= i\phi\psi, \qquad \psi_{[2]}= -\phi^2\psi, \qquad
\psi_{[3]}=-i\phi^3\psi.$$
This is in perfect agreement with the perturbative expansion of
the ``dressed Dirac field''
$\psi_{qc}= e^{iq\phi(c)}\cdot \psi$ as discussed in \cite{MRS3} for the
QED coupling to massless photons, where a version of \eref{XLV} is
used to construct the interacting fields. In contrast to the massless
case where $\psi_{qc}$ can be defined as a string-localized field,
the massive Wick exponential $e^{ig\phi(c)}$ is a ``Jaffe field'' of very poor
localization properties \cite{Ja}.

\section{Comparison with other approaches}
\label{s:comp}

We compile here the various approaches to the Abelian Higgs model, not least because the same symbols tend to stand for
different objects in different settings. See for this \sref{s:syn}.

\subsection{Spontaneous symmetry breaking in unitary gauge}
\label{s:SSB}

In the textbook narrative of the Higgs mechanism, one starts classically
with the minimal coupling
of a charged scalar Higgs field $\Psi$ with the 
potential $U(\Psi^*\Psi)$ to a massless vector field $B$ with field strength
$G=\pa\wedge B$:
\bea{LSSB}L=-\frac14 G_{\mu\nu}G^{\mu\nu}  + (D_\mu\Psi)^*D^\mu \Psi
-U(\Psi^*\Psi), \qquad U(\Psi^*\Psi)= \kappa \Big(\Psi^*\Psi-\frac
{v^2}2\Big)^2.
\eea
Parameterizing
$$\Psi(x) = \frac1{\sqrt2}(v+H(x))e^{i\chi(x)/v},$$
one can ``gauge away'' the Goldstone mode $\chi(x)$. One then
writes the resulting Lagrangian in terms of $B$ and $H$. After suitable identification of
the parameters ($m^2:=g^2v^2$ where $g$ is the gauge coupling
constant, and $m_H^2:=2\kappa v^2$), one arrives at
\bea{Lfree}L=-\frac14 G_{\mu\nu}G^{\mu\nu} + \frac{m^2}2B_\mu
  B^\mu  + \frac
12\pa_\mu H\pa^\mu H -\frac{m_H^2}2H^2 +L\P, \eea
which contains the free massive Proca and 
Higgs Lagrangians along with the interaction density 
\bea{LProca}
L\P = mg \Big(B^2H - \frac{m_H^2}{2m^2}H^3\Big) +
\frac{g^2}{2} \Big(B^2 H^2- \frac{m_H^2}{4m^2}H^4\Big).
\eea
The Higgs mass term in \eref{Lfree} and the cubic and quartic terms in \eref{LProca}
together constitute the Higgs potential \eref{VH}. As a
quantum interaction, \eref{LProca} is non-renormalizable.

By regarding the unitary gauge as a limiting case at tree
level\footnote{This is not true for loop corrections \cite{Wu}.} of the
renormalizable $R_\xi$ gauges in indefinite metric, and exploiting the unbroken gauge
invariance to establish the necessary Ward identities, it is
concluded that the theory is renormalizable and unitary.

\subsection{BRST approach}
\label{s:BRST}

In \cite{ADS} and \cite[Chap.~4.1]{Scha}, the Abelian Higgs model is
constructed
without spontaneous symmetry breaking. The method of securing BRST
invariance of the S-matrix is similar to our $L$-$Q$-pair method:
one recursively fixes induced interaction terms to cancel obstructions
in each order.

One starts on the indefinite Fock space (Krein space) of the massive
vector potential $A\K$ in the Feynman gauge, the Higgs
field $H$, the ghost
fields $u$, $\wt u$, and the independent positive-definite scalar Stückelberg field $\Phi$ of
the same mass as $A\K$ \cite{RR}. The cubic interaction is given by the
power-counting renormalizable interaction density\footnote{\label{fn}In \cite{ADS,DGSV,Scha},
the Higgs field $H$ is denoted by $\phi$ or $\varphi$. We reserve
$\phi$ for the escort field.}
\bea{LBRST} L^{\rm BRST} = mg\Big(A\K(A\K H + \frac1m \Phi \lrpa
H) + u\wt uH - \frac{m_H^2}{2m^2}\Phi^2 H + aH^3\Big).
\eea
The nilpotent BRST transformation $s$ is implemented by the graded
commutator with a nilpotent free BRST
operator $Q$, whose cohomology $\HH=\mathrm{Ker}(Q)/\mathrm{Ran}(Q)$ is the positive-definite
physical Hilbert space of the free theory. The BRST variations $s(X) =
i[Q,X]_\pm$ are 
$$s(A\K_\mu) = -\pa_\mu u, \quad s(\Phi)=-mu,\quad s(u)=0,\quad s(\wt u) =
\pa_\mu A\K{}^\mu+m\Phi, \quad s(H)=0.$$
The interaction \eref{LBRST} is distinguished by the property that its BRST
variation is a total derivative:
$$s \big(L^{\rm BRST}\big) = \pa_\mu\big((mA\K{}^\mu H+\Phi\lrpa{}^\mu H)u\big).$$
The nontrivial condition to secure Hilbert space
positivity of the interacting theory is that this feature must
persist in higher orders of perturbation theory for the S-matrix. 
This condition in second order requires to add quartic terms
$$\frac{g^2}{2} \Big(A\K{}^2 H^2+
A\K{}^2\Phi^2-\frac{m_H^2}{4m^2}\Phi^4+\big(3a+\frac{m_H^2}{m^2}\big)\Phi^2H^2
+ bH^4\Big)$$
to the interaction density, in
order to cancel obstructions. The cubic and quartic coefficients of
the Higgs potential are fixed in third order. The result are the values \eref{ab}. 

The field
\bea{Bemb} B_\mu:=A_\mu\K-m\inv \pa_\mu \Phi\eea
is BRST-invariant. Its two-point function coincides with the positive-definite Proca two-point function \eref{BB}. However, $\pa_\mu B^\mu=\pa_\mu A^\mu + m\Phi\neq
0$ on the Fock space. But this quantity is a null field in the range of
$s$, hence it vanishes on
the BRST Hilbert space $\HH$. Thus, \eref{Bemb} on $\HH$ is the Proca
field. Moreover, on $\HH$ also $u$ vanishes, and the interaction
density \eref{LBRST} coincides with the cubic part of \eref{LProca} up
to a total derivative.

\subsection{String-localized $L$-$Q$-pair approach (on the Hilbert space)}
\label{s:LQ}

The $L$-$Q$-pair approach was exhibited at length in
\sref{s:ahmLQ}. Apart from the values of the parameters $a$ and $b$,
its result $L[\chi;c]$ cannot be directly compared to point-localized
approaches. But it can be asserted that 
the initial $L$-$Q$-pair \eref{LQini} defines a string-independent
S-matrix iff the parameters $a$ and $b$ in $L_1$ and $L_2$ take the
values \eref{ab}.

\subsection{String-localized $L$-$V$-pair approach  (on the Hilbert space)}
\label{s:LV}

The $L$-$V$-pair approach was exhibited at length in
\sref{s:ahmLV}. Here, the assertion is that for the initial $L$-$V$ pair
\eref{LVini}, the  S-matrix in the left-hand side of \eref{SILV} coincides with (and
actually defines in the adiabatic limit) the string-independent S-matrix 
on the right-hand side.

Collecting the pieces $\Lp_n$ in \eref{Lp} as computed in
\sref{s:ahmLV} with the choice \eref{ccH} of renormalization parameters, one gets
\bea{Lptot}\Lp = mg\cdot\big(B^2H - \frac{m_H^2}{2m^2}H^3\big) + \frac{g^2}2\cdot\big(
-3B^2H^2- \frac{m_H^2}{4m^2}H^4  \big) + \frac{g^3}6\cdot\frac{12}m
B^2H^3 + \dots,\eea
It contains the interaction part of the Higgs potential \eref{VHab}
whose coefficients are the same as in all other
approaches; plus possibly further coupling terms like $g^n\cdot
  B^2H^n$ ($n>3$). The latter will be discussed in \sref{s:syn}.

  The string-localized interaction density $L[c]$ is also
    defined on the Krein space of \sref{s:BRST}, where $B$ is given by
  \eref{Bemb} and $A(c)$ and $\phi(c)$ are defined by \eref{ABphi} and
  \eref{phiBe} (smeared with $c(e)$). It then holds, for any $c$
  \bea{sphi} s(\phi(c))=u.\eea
The cubic SQFT interaction \eref{L1} (with
  $a'=\alpha_i=0$) differs from \eref{LBRST} by 
  $$L_1(c) = L_1^{\rm BRST} - s\big((m\phi(c)-\Phi)\wt u H\big) +
  \pa_\mu\big(B^\mu(m\phi(c)-\Phi)H + \frac 1{2m}
(m^2\phi(c)^2-\Phi^2)\pa^\mu H\big) .$$
Up to the term in $\mathrm{Ran}(s)$ which can be included in $L^{\rm
  BRST}$ at no expense, this is another
instance of the example mentioned in the beginning of \sref{s:LVLQ}.

\newpage

\subsection{Krein space $L$-$V$-pair approach (point-localized)}
\label{s:Krein}
One may also work with a point-localized $L$-$V$ pair in the Krein space of the
massive vector potential $A\K$ and the Higgs field, without the ghost and Stückelberg fields.
The Proca field $B$ is embedded into the Krein space as
\bea{emb}B_\mu: = m^{-2} \pa^\nu G\K_{\mu\nu} = A\K_\mu-  \pa_\mu \phi\K\eea
where $G\K:= \pa \wedge A\K$ is the field strength, and $\phi\K :=
-m^{-2} (\pa A\K)$. We refer to the subspace generated from the
  Krein vacuum
by $B_\mu$ and $H$ as the ``embedded Hilbert space''.

Then one has an $L$-$V$ pair
\bea{LVK}L\K_1 = L\P_1+\pa V\K_1\eea
with
\bea{LK1}
L\P_1&=&m\cdot\big(B^2H +aH^3\big),\\ \notag
L\K_1 &=& m\cdot\Big(A\K B H + A\K\phi\K \pa
H - \frac{m_H^2}{2}\phi\K{}^2 H + aH^3\Big), \\ \notag
V\K_1 &=& m\cdot\Big(B\phi\K H + \frac 12 \phi\K{}^2\pa H\Big).\eea
We want to use this pair as the starting point of a recursion as
in \sref{s:PSI-imp} to
reformulate a non-renormalizable point-localized interaction of
the Proca and Higgs fields on the embedded Hilbert space, as a
renormalizable point-localized interaction on the Krein space.  The
PSI in this case is replaced by the principle of Hilbert space
positivity, i.e., the right-hand side of the analogue
$$Te^{i(L\K[\chi;c] + V\K[\chi;c]) }= Te^{iL\P[\chi]}$$
of the identity \eref{SILV} should be defined on the embedded Hilbert space.

We then proceed as in \sref{s:PSI-imp} and recursively determine the
higher-order densities with the specification that $L\K_n$ are
renormalizable are $L\P_n$ are defined on the embedded Hilbert space. This would secure a
positive-definite renormalizable theory in the adiabatic limit.

The triple $L\P_1$, $L\K_1$, $V\K_1$ is identical with the
triple $\Lp_1$, $L_1(c)$, $V_1(c)$ in \sref{s:ahmLV} with $A(c)$ replaced
by $A\K$ and $\phi(c)$ replaced by $\phi\K$. However, the
two-point obstructions are different, due to the different scaling
degrees of the two-point functions and different linear relations
among the fields and their derivatives, see in \aref{a:krein-2pt}. 
The two-point obstructions \eref{OH} in the Higgs sector are
unchanged, those in the vector boson sector are
 \bea{OBK}O_\mu(A\K{}^\mu;B'_\nu)= O_\mu(B^\mu;A\K_\nu{}')=
O_\mu(B^\mu;B'_\nu)&=&-i(1+c_B)\cdot m^{-2}\partial_\nu\delta(x-x'), \notag \\ 
O_\mu(A\K{}^\mu;\phi\K{}') &=& -im^{-2}\delta(x-x'), \notag \\
O_\mu(\phi\K;A_\nu\K{}')&=&-ic_B\cdot
m^{-2}\eta_{\mu\nu}\delta(x-x'), \notag \\ 
O_\mu(A\K{}^\mu;A_\nu\K{}') =
O_\mu(B^\mu;\phi\K{}')=O_\mu(\phi\K;B'_\nu) = O_\mu(\phi\K;\phi\K{}') &=&0.\eea
With these, one computes
the second-order obstruction \eref{O2O} of the
$L$-$V$ pair $L\K_1=L\P_1+\pa V\K_1$. One finds that it can be
  cancelled with
\bea{LK2} L\P_2 \!\!&=&\!\! (1+4c_B)\cdot B^2H^2+bH^4  ,\\ \notag
L\K_2\!\!&=&\!\!
(A\K{}^2+3c_BA\K B)H^2+m^2\Big((3a+\frac{m_H^2}{m^2}+c_B)\phi\K{}^2H^2-\frac{m_H^2}4\phi\K{}^4
+(1+c_H)A\K{}^2\phi\K{}^2\Big) +bH^4, \\ \notag
V\K_2 \!\!&=&\!\! \big(A\K-(1-c_B)\cdot B\big)\phi\K H^2 +\frac{m^2}6B\phi\K{}^3  + (1+c_H)\cdot \frac
{m^2}2A\K\phi\K{}^3 +c_B\cdot \phi\K{}^2 H\pa H, \\ \notag
W\K_2\!\!&=&\!\!  (1+c_H)\cdot \frac {m^2}4\phi\K{}^4 + (1+c_B)\cdot \phi\K{}^2H^2 
\eea
with a free coefficient $b$ of the the quartic
part of $V(H)$. However, the term $A\K BH^2$ in $L_2\K$ has dimension 5 and is not
  renormalizable. One therefore has to choose $c_B=0$. 
Quite amazingly, precisely with this choice the expressions
\eref{LK2} are identical with \eref{L2} (with the replacement of
string-localized fields by Krein fields).

With $c_B=0$, the complete computation of the third-order obstruction as in
\eref{O3O} is more contrived
because of the derivatives of $\delta$-functions. We have used
\eref{IO3O} to compute it up to derivatives as in \eref{canc3}. It
turns out that the third-order obstruction cannot be
cancelled by third-order densities with $L\P_3$ positive-definite and
$L\K_3$ renormalizable, for any value of $c_H$. Thus, this approach
fails in third order.

\subsection{String-localized $L$-$V$-pair approach in Krein space}
\label{s:sKLV}
For the sake of completeness, we report yet another $L$-$V$ pair,
which reformulates the point-localized Krein space interaction as in
\sref{s:Krein} as the renormalizable string-localized Hilbert space 
interaction as in \sref{s:ahmLV}. Unlike in \sref{s:Krein}, non-renormalizable
  higher-order terms $L\K_n$ are admitted in the Krein space  interaction.
We thus 
want to establish the identity
\bea{sKLV}Te^{i(\wt L[\chi;c] + \wt V\!\circ\pa[\chi;c])} 
\stackrel!=Te^{i\wt L\K[\chi] }\eea
with the initial $L$-$V$ pair 
$$\wt L_1=\wt L\K_1+\pa \wt V_1,$$
where $\wt L_1= L_1$ as in \eref{LVini} and $\wt L\K_1=L\K_1$ as in
\eref{LK1}, thus (because $L\P_1$ is identical with $\Lp_1$ embedded
into the Krein space)
$$\wt V_1 = V_1-V\K_1.$$
Along
with the Proca field \eref{emb}, also the string-localized fields are
embedded into the Krein space via \eref{ABphi} and \eref{phiBe}, and it holds
\bea{2BAphi} B=A-\pa\phi = A\K-\pa \phi\K.\eea
The two-point obstructions among the Hilbert space fields and among
the Krein space fields are as before. One also needs mixed 
two-point obstructions, which turn out to be
$$O_\mu(\phi;\phi\K{}') = O_\mu(A^\mu;\phi\K{}') =
O_\mu(\phi\K;\phi')=O_\mu(\phi\K;A'_\nu)=0,$$
$$O_\mu(\phi;A\K_\nu{}') = O_\mu(\phi;B'_\nu)= -ic_B\cdot m^{-2}\eta_{\mu\nu}\delta_{xx'},  \quad
O_\mu(A\K{}^\mu;\phi')=O_\mu(B^\mu;\phi')=-im^{-2}\delta_{xx'},$$
$$\quad O_\mu(A^\mu;A\K_\nu{}') = O_\mu(A^\mu;B'_\nu) = -ie_\nu I_e\delta_{xx'},\quad
O_\mu(A\K{}^\mu;A'_\nu)=O_\mu(B^\mu;A'_\nu)=0.\qquad\,$$
Notice that $O_\mu(A^\mu;A\K_\nu{}')$ is string-localized.

Because in this approach, the right-hand side of \eref{sKLV} is not
required to be renormalizable, terms like $A\K BH^2$ are admitted in $\wt
L\K_2$ (in contrast to $L\K_2$ in \sref{s:Krein}.) The second-order obstruction can be cancelled by $\wt
L_2-\pa\wt V_2-\wt L\K_2$ for arbitrary values of $c_B$ and $c_H$, but
$\wt V_2$ contains terms involving the string-localized field $A$ with
coefficients $(1+c_B)$ or $(1+c_H)$. As in \sref{s:ahmLV}, such terms
would produce string-localized $\delta$-functions in the third-order obstruction,
which cannot be cancelled. Therefore, we have to choose again $c_B=c_H=-1$. With this choice,  
\bea{sKL2}
\wt L_2&=& m^2\Big((3a+\frac{m_H^2}{m^2})\phi^2H^2 -
\frac{m_H^2}{4}\phi^4\Big), \\ \notag
\wt L\K_2&=& A\K(A\K- 3B)H^2 +m^2\Big(
\frac12(B-A\K)B\phi\K{}^2 + (3a+\frac{m_H^2}{m^2}-1)\phi\K{}^2H^2 -
\frac{m_H^2}{4}\phi\K{}^4\Big),\\ \notag
\wt V_2&=& (B-A\K)\phi\K H^2-B\phi H^2+\frac{m^2}6B\phi^3 +
\frac{m^2}2B(\phi\K-\phi)\phi\phi\K + (\phi\K-\phi)\phi\K H\pa H.\eea
We have then computed the
third-order obstruction using \eref{O3simp}.\footnote{It contains 39 terms.
Although the explicit expressions are of little interest, we just report the final $\wt L\K_3$: 
$$L\K_3= m^{-1}(2A\K{}^2-11A^K B)H^3
+ m\big((3A\K{}^2+\frac32 A\K B+3B^2)\phi\K{}^2H
+ (2-3\frac{m_H^2}{m^2})\phi\K{}^2H^3\big)
+m^3(1-\frac{m_H^2}{4m^2})\phi\K{}^4H.$$
} All its ``mixed
terms'' (products of string-localized and Krein fields) and
string-localized terms can be
cancelled by derivatives $\pa \wt V_3$, except precisely the same last two terms as
in \eref{O3final}. Because $\wt L_3$ must
vanish by renormalizability of the left-hand side of \eref{sKLV}, this means that the
third-order obstruction can be cancelled if
and only if the parameters $a$ and $b$ take the values \eref{ab} of the Higgs
potential.

\subsection{Synopsis}
\label{s:syn}

In all approaches \sref{s:BRST}--\sref{s:LV}, one has the
renormalizable interaction density of the form
$$L_1=m\Big(ABH + A \phi \pa H -\frac12m_H^2 \phi^2H+ a H^3\Big)$$
(plus ghost terms in the BRST setting), and a relation of the form
$$B_\mu = A_\mu - \pa_\mu \phi.$$
However, not only the meanings of the symbols $A_\mu$ and $\phi$ are very
different, but also their correlations, hence propagators and
obstructions in perturbation theory. This explains the different induced
quartic and higher interaction densities found in the various approaches. 

In the string-localized approach of the main body of the paper (\sref{s:ahmLV}), $A=A(c)$ is a string-localized potential
and $\phi=\phi(c)$ its string-localized escort field (an integral over
the Proca field), both defined on
the physical Hilbert space of the Proca field. In BRST (\sref{s:BRST}),  $A=A\K$ is the Feynman
gauge vector potential and $\phi$ is (up to the
factor $m$) the independent positive-definite Stückelberg field $\Phi$
with
$\erw{A\K\Phi}=0$. In the Krein space $L$-$V$-pair approach (\sref{s:Krein}), $A=A\K$
as in BRST, but $\phi=\phi\K=-m^{-2}(\pa A\K)$ is a derivative of the
former and negative-definite. Finally, in \sref{s:sKLV}, we
have two sets of fields $A(c), \phi(c)$ and $A\K,\phi\K$, related to
each other by the two representations \eref{2BAphi} of the Proca field.

In the BRST approach, $\pa A+m\Phi$ is in the range of the BRST
transformation, hence it vanishes on the physical Hilbert space. Thus,
the positive-definite Stückelberg field is ``identified'' with the negative-definite
$m\phi\K=-m\inv (\pa A\K)$ of the Krein space approach. This is of course only possible because their difference
is a null field, which is zero in the BRST quotient space.

It is remarkable that, although the obstructions appearing in
perturbation theory are different, they can be cancelled in the BRST
and the string-localized approaches, and the fixing of the
coefficients $a$, $b$ of the Higgs potential in third order gives the same
values in all of them. The fact that the cancellation is not possible
in the point-localized Krein space approach (without a Stückelberg field) shows
that it is by no means automatic that identities like \eref{SILV} can be
recursively fulfilled. Rather, there must be some
hidden features of the model whose general nature is not transparent to us. Apparently, string-localization provides the necessary
flexibility that in the BRST and the string-localized Krein
space approaches is provided by the blowing-up of
the field content.

The $L$-$Q$-pair approach does not allow to compute $\Lp_n$. For $L_n$,
it gives compatible results with the $L$-$V$-pair approach.

Having established, by virtue of the identity \eref{SILV}, the equivalence between the renormalizable
string-localized interaction $L(e)$ with the non-renormalizable
point-localized interaction $\Lp$, we should
ask whether the latter is equivalent to the interaction $L\P$ as in \sref{s:SSB}. $\Lp$ in
\eref{Lptot} and $L\P$ in \eref{LProca}
differ by the coefficient of the second-order term
$B^2H^2$ and by a new third-order term $B^2H^3$ (and possibly higher-order
terms $B^2H^n$).\footnote{Also the BRST approach in \sref{s:BRST} produces a different quartic
interaction, except for the Higgs potential.} On the other hand, the
former uses the renormalization of the Proca propagator with $c_B=-1$,
that was necessary in order to eliminate string-localized
obstructions in third order. The latter uses the 
kinematical choice $c_B =0$. We shall now give evidence for
the presumed equivalence.

This apparent discrepancy is a variation of
the familiar observation in scalar QED, that the
renormalization (by adding a multiple of the
$\delta$-function) of a propagator connecting two cubic vertices,
just amounts to another quartic vertex. That the renormalization of
the propagator $\erw{T\pa_\mu\varphi\pa_\nu\varphi}$ can be traded for
the coefficient of the 
quartic vertex $A^2\varphi^*\varphi$ of scalar QED, has been proven in all orders in
various settings \cite{DKS,DPR,Ti}. Requiring gauge invariance of the total
Lagrangian would select the kinematical propagator. But causal
perturbation theory does not need gauge invariance and can be done
with an arbitrary renormalization. The result is equivalent up to a
renormalization group transformation interpolating between both values.

The case at hand, with many vertices $B^2H^n$ and the Higgs
  self-couplings, is more contrived than scalar QED.  Yet, for the
  tree-level scattering amplitudes for processes $$\hbox{2 vector
  bosons} \to \hbox{$n$ Higgs},$$
when computed 
with \eref{Lptot} in the $L$-$V$-pair approach ($c_B=-1$) and with
\eref{LProca} in the unitary gauge ($c_B=0$), we have
verified the match for $n=2$ and for $n=3$, as follows.

For the comparison of different values of the renormalization
parameter, \eref{TBB} can be written graphically (solid lines $=B$, $n_1+n_2$ broken lines $=H$) as
\bea{shrink}\includegraphics[width=60mm]{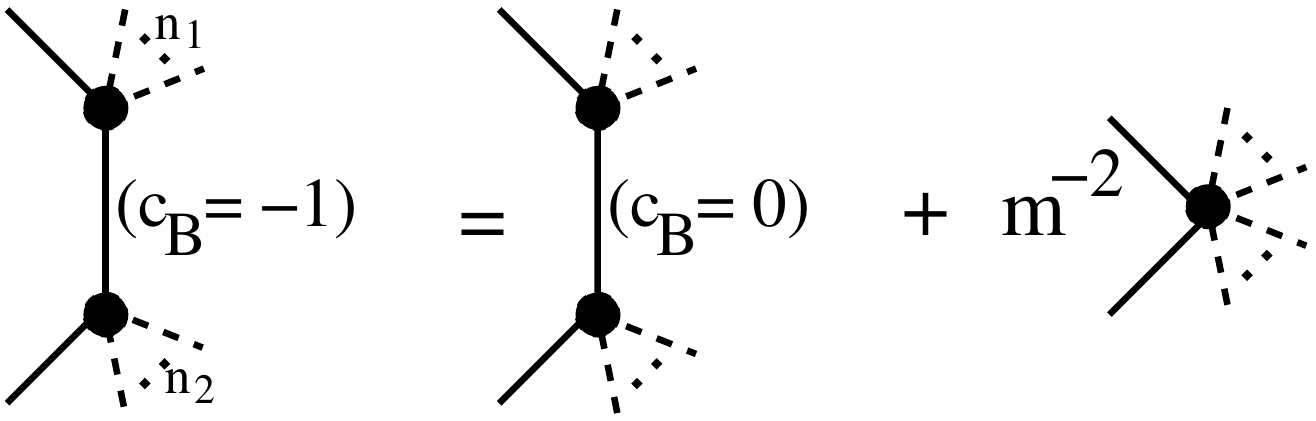}.\eea
This immediately implies the match between \eref{Lptot} and
\eref{LProca} for $n=2$ in order $g^2$:
$$\frac{-3}{2}\cdot 4 \cdot\raisebox{-4mm}{\includegraphics[height=10mm]{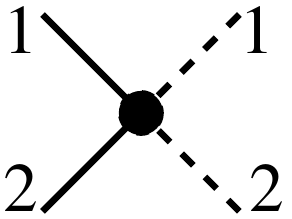}} + \frac {m^2}2\cdot 8\cdot
\left[\raisebox{-8mm}{\includegraphics[height=18mm]{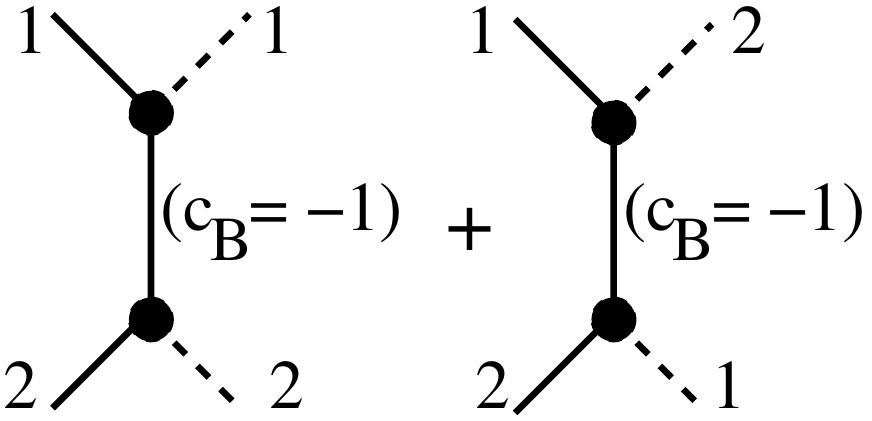}}\right]= \frac{1}{2}\cdot 4 \cdot\raisebox{-4mm}{\includegraphics[height=10mm]{match.pdf}} + \frac {m^2}2\cdot 8\cdot
\left[\raisebox{-8mm}{\includegraphics[height=18mm]{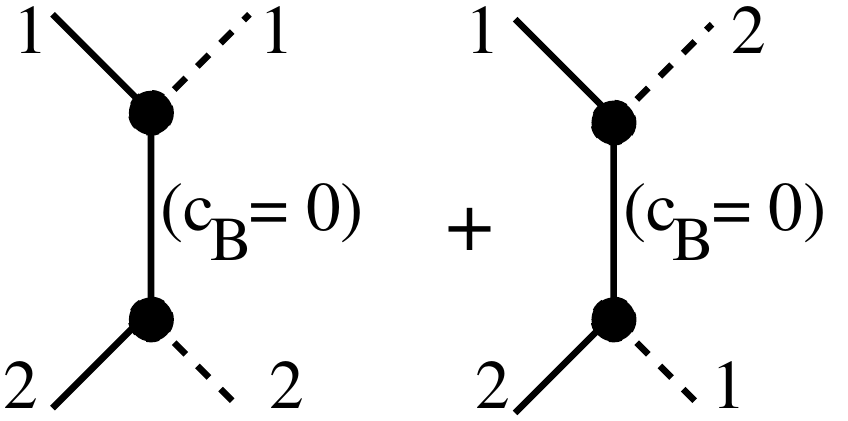}}\right] $$
(where the factors $4$ and $8$ are the counting factors for equivalent contractions).
Note that the coupling
constants for the $B^2H^2$-term differ by the factor of $-3$. The
difference is made up by the contributions from the second diagram on
the left-hand side to the first diagram on the right-hand side, due to \eref{shrink}.
By the same method, we have verified the match also in the case $n=3$ (seven diagrams with
permutations, of which two diagrams need not be considered because
they do not contain differing coupling constants or renormalized
propagators), -- thus justifying the presence of the term $B^2H^3$ in
$\Lp_3$, and confirming the value of its coefficient.

In view of this evidence for the equivalence between $L\P$ in
  \eref{LProca} with $c_B=0$ and $\Lp$ in \eref{Lptot} with $c_B=-1$,
it would be most rewarding to find again a renormalization 
  group transformation interpolating between them.

All approaches discussed here may be regarded as attempts to
  define a renormalization of a power-counting non-renormalizable
  interaction. They do not differ in their physical
predicitions, but in the way how (and whether) fundamental principles of quantum
field theory are implemented. What stands out is the universality
  of the Higgs potential, that is the same in all consistent
  approaches. Its universal shape is recognized to be an intrinsic consistency
  condition, rather than an input to trigger a spontaneous breaking of
  gauge symmetry.

\section{Discussion}
\label{s:disc}

The ubiquitous clashes between Hilbert space, causality, and
renormalizability are worrying us since the early days of QFT. 
The $L$-$V$-pair formalism developped in this paper allows to establish equivalences between 
formulations of QFT models, in which complementary subsets of these fundamental principles are fulfilled, 
such that, by the very equivalence, all of them hold simultaneously --
but possibly not in 
any single formulation. Even more, it allows to fix physical
parameters (coupling constants) as consistency conditions for the
equivalences to hold.

In particular, by providing the necessary $L$-$V$ pairs, 
string-localized QFT can be employed with various benefits. In the present paper, we
have considered an instance where it can be used to ``renormalize the
unrenormalizable'', provided certain parameters are appropriately fixed
to secure consistency of the method. The physical manifestation of these parameters 
is the Higgs potential. It owes its universal shape to the fundamental
principles of Hilbert space positivity and locality, rather than an
aesthetic but positivity-violating gauge principle.

In the same way, SQFT has been used earlier to explain the ``gauge theory pattern''
of massive vector boson self-couplings \cite{GGM}, and the chirality of the weak
interaction \cite{GMV}.

In a very different way, it has been used to explore the infrared
structure of QED. Here, the logarithmic infrared divergence of the
string integration defining the escort field becomes instrumental for
a new understanding of the superselection structure of QED \cite{MRS3}, in which
the string smearing function describes the ``shape of the photon
cloud'' of charged states \cite{MRS2}. SQFT allows to construct string-dependent
charged fields: the Principle of String Independence holds for the S-matrix
 in the neutral sector and for observable fields, but string
 dependence of charged states and unobservable
 fields becomes a physical feature (the ``shape of photon
 clouds'').

 QED is in fact the prototype of an SQFT, which has no
 second-order obstructions and hence no higher-order interactions
 $L_n$ ($n>1$). This property distinguishes QED from the model treated
 in this work, and allows
 a non-perturbative construction leading ``halfways'' to QED \cite{MRS3}.

It is natural to consider an SQFT treatment of QCD. Massless Yang-Mills
theory is an instance where an $L$-$Q$ pair
\bea{YMLQ} L_1(c) &=& f_{abc} A^a_\mu(c) A^b_\nu(c)F^{c\mu\nu}  , \\
\notag Q_1^\mu &=& 2f_{abc} w^a A^b_\nu(c) F^{c\mu\nu} , \eea
with $f_{abc}$ completely antisymmetric and $w^a = \delta_c A^a(c)$, 
exists on the Wigner
Hilbert space of the free field strengths $F^a_{\mu\nu}$, but no $L$-$V$ pair  \cite{G2}.
(Notice that the same expression \eref{YMLQ} on the Krein space, as in
\sref{s:Krein}, is not even an $L$-$Q$ pair, because $\pa_\mu
A^{a{\rm K}\mu}\neq 0$.) 

The case of QCD still remains to be worked out. The $L$-$Q$ pair \eref{YMLQ} has second-order
obstructions, so that there is no immediate analogue of the non-perturbative
construction that gives rise to the infrared superselection structure
of QED. Instead, it is expected that no color-charged states can be
constructed at all, which would be a new mechanism to explain
confinement.

The use of string-localized quantum fields in the interaction gives us
occasion to comment on the fact (underlying causal
perturbation theory also in the point-localized case): {\em Interaction
does not need a free Lagrangian.}
This is advantageous, because ``canonical quantization'' based on free
Lagrangians is beset with difficulties.
The zero-component of the Maxwell four-potential has no canonically
conjugate momentum: one needs a ``gauge-fixing term'' to
cure this problem, and one needs another cure (the Gupta-Bleuler
condition) to make the first cure ineffective for the dynamics. Why is the
classically purely auxiliary four-potential treated as  
fundamental in the first place, and not the observable Maxwell
field tensor? For massive tensor fields of higher spin,  ``free
Lagrangians''  need a host of auxiliary fields to implement constraints \cite{Fro}. For
spinor fields, {\em anti-}commutation relations have no a
priori ``canonical'' justification: they are needed to reconcile covariance with
Hilbert space positivity after the quantization has been performed at
the one-particle level, and Dirac's theory to deal with first-class
constraints is needed to save the idea of canonical quantization with
a free Lagrangian that is linear in the momenta.

Weinberg \cite{Wein} has shown how one can bypass all these
pains. Given a unitary one-particle representation of the Poincar\'e
group as classified by Wigner \cite{Wig}, one directly constructs free fields on the Fock space (i.e.,
a Hilbert space) over the
one-particle space. Their interaction is described by the {\em interaction}
density $L\equiv L_{\rm int}$ alone. There is no reference to an
interacting equation of motion (which in the literal sense does not
exist in QFT). The interacting quantum field is constructed
perturbatively by ``causal perturbation theory'' due to Glaser and
Epstein \cite{EG}, who turned Bogoliubov's somewhat heuristic formula
\cite{Bog} into a rigorous working scheme.

A side-message of Weinberg's construction is that quantum fields
associated with a given particle are by no means unique: the
intertwiner condition on the coefficients of creation and annihilation
operators has many solutions. The resulting fields are all defined on
the same Fock space and create the same particle states. This is
trivially true for
derivatives of a given field, and derivatives are the only operations that respect causal (anti-)commutativity. But if one is willing to relax localization (e.g., in
order to tame the UV singularity of the propagator), then 
more flexibility is gained with string-localized fields. Even the NoGo result against local fields 
for infinite-spin particles \cite{Yng} can be overcome \cite{MSY}
without the need to sacrifice the Hilbert space.

To conclude: There exist several different
  but equivalent ways to set up the perturbation theory of the same
  QFT model.  The setups may be competitive in which fundamental
  principles they respect manifestly, and which ones have to be concluded
  indirectly. In the case of the Abelian Higgs model, SQFT seems to be
closest to the ``best of all worlds'' in which Hilbert space
positivity, covariance, locality and renormalizability are all
satisfied at the same time and at every step. (String-localization is
a very mild relaxation of locality for non-observable fields.) By avoiding unphysical field
degrees of freedom, it is also the most economic one.
In addition, we have stressed that the precise shape of the Higgs potential
is determined by internal consistency with fundamental principles, without invoking the usual gauge theoretical arguments.

\bigskip

{\bf Acknowledgments:} JM was partially supported by the Emmy Noether
grant DY107/2-2 of the Deutsche Forschungsgemeinschaft.

\appendix

\section{Second and third-order SI conditions}
\label{a:SI}

For the proofs of \pref{p:O2LV} and \pref{p:O3LV} it is 
immaterial that the densities $L_n(x,c)$ are string-localized and
$\Lp_n(x)$ point-localized. We write more generally just $L_n$ and $K_n$ instead,
so as to cover also the $L$-$V$ pairs to be discussed in \sref{s:Krein} and \sref{s:sKLV}.

{\em Proof of \pref{p:O2LV}:}
The expansion of \eref{SILV} in second order reads 
$$\frac{i^2}2\int dx\,dx'\,
\big(\chi\chi'\cdot T[L_1L_1']+\pa_\mu\chi\chi'\cdot T[V^\mu_1L'_1]+\chi\pa'_\nu\chi'\cdot T[L_1V_1'^\nu]+\pa_\mu\chi\pa'_\nu\chi'\cdot 
T[V^\mu_1V'^\nu_1]\big)-$$$$ +\frac{i}2 \int dx\, (\chi^2\cdot L_2+
\pa_\mu\chi^2\cdot V_2^\mu + \pa_\mu\chi\pa_\nu\chi \cdot W_2^{\mu\nu})
\stackrel!=
\frac{i^2}2\int
dx\,dx'\, \chi\chi' \cdot T[K_1K'_1] +\frac{i}2 \int dx\, \chi\cdot K_2.
$$
We insert the initial $L$-$V$ pair relation
$L_1=K_1+\pa_\mu V_1^\mu$, and integrate by parts. After the
obvious cancellations, this becomes the determining
condition for $L_2,K_2,V_2,W_2$
$$\int dx\,dx'\,\chi\chi'\cdot \Big[(O^{(2)}(x,x')
-i\delta(x-x')(L_2-K_2-\pa_\mu
V_2^\mu) -\pa_\mu\pa'_\nu\big(i\delta(x-x')W_2^{\mu\nu}\big)
\Big]\stackrel!=0,$$
where
$$O^{(2)}(x,x')=[T,\pa_\mu]
V_1^\mu K_1{}'+[T,\pa'_\mu]K_1
V_1'^\mu+[[T,\pa_\mu],\pa'_\nu]  V_1^\mu
V_1'^\nu. $$
With the notation \eref{OVdef}, this is \eref{O2O}. \qed

{\em Proof of \pref{p:O3LV}:} 
Expanding \eref{SILV} in third order, 
eliminating $L_1$ by the first-order condition, and integrating by
parts, we get
$$\int dx\, dx'\, dx'' \, \chi\chi'\chi'' \big[O^{(3)}(x,x',x'')-\delta_{xx'x''}(L_3-K_3-\pa
V_3) - \SS_3\big(\pa\pa'[\delta_{xx'x''}W_3]\big)\big] \stackrel!=0,$$
where
\bea{O3}
O^{(3)}(x,x',x'')\hspace{-2mm}&=&\hspace{-2mm}\SS_3\Big(3[T,\pa]V_1K_1'K_1''
+ 3[[T,\pa],\pa']V_1V_1'K_1'' + [[[T,\pa],\pa'],\pa'']
V_1V_1'V_1'' -  \\
&-& \notag\hspace{-2mm}3i\delta_{x'x''}\big([T,\pa]V_1L_2' +[T,\pa']V_2'K_1-\pa'[T,\pa]V_1V_2'\big)  -
\\ &-& \hspace{-2mm}3i\delta_{x'x''} TK_1(L_2'-K_2'-\pa'V_2')-3i\pa'\pa''\big(\delta_{x'x''}([T,\pa]V_1W_2' +
TK_1W_2')\big)\Big) . \qquad\notag \eea
The subsequent \lref{l:Lb} is the tree-level version of the ``Master Ward Identity'' of
\cite[Sect.~2.4]{BD}, which the authors postulate to hold as a natural renormalization
condition in all loop orders. It will allow substantial systematic
cancellations in \eref{O3}.

\blemma{l:Lb}
For Wick polynomials  $Y$ and $X_i$, let
\bea{OVXY}
O_\mu(Y;X_1,\dots,X_n) := [T,\pa_\mu] Y(x) X_1(x_1),\dots,X_n(x_n)\tree.
\eea
It holds
\bea{OVXYtot}O_Y(\wick{X_1},\dots,\wick{X_n}) = \sum_{i=1}^nT\big(\wick{O_Y(X_i)}\wick{X_1}\dots
\wick{\big\slash \!\!\!\!X_i}\dots \wick{X_n}\big)\tree.
\eea
 \elemma
 {\em Proof:}  We insert $\pa {Y} =
 \sum_\varphi{\frac{\pa Y}{\pa\varphi}\pa\varphi}$ (as Wick polynomials) in 
$T\big(\pa {Y} {X_1}\dots {X_n}\big)$. In the Wick expansion, the terms in which $\pa\varphi$ is
not contracted, cancel against the corresponding terms in the Wick
expansion of $\pa T\big({Y}
{X_1}\dots {X_n}\big)$ in which the derivative hits a noncontracted factor
of $Y$.

The terms in $T\big(\pa {Y} {X_1}\dots {X_n}\big)$ in which $\pa\varphi$ is
contracted with one of the fields $X_i$, can be written as
$$\sum_{\varphi,\chi_i}\erw{T\pa_\mu\varphi\,\chi_i}\cdot T\big(\wick{\frac{\pa Y}{\pa\varphi}\frac{\pa X_i}{\pa\chi_i}} \wick{X_1}\dots
\wick{\big\slash \!\!\!\!X_i}\dots \wick{X_n}\big)\tree.$$
Note that at tree level, there are no further contractions between $\frac{\pa
  Y}{\pa\varphi}$ and $\frac{\pa X_i}{\pa\chi}$, so the latter appear in a
single Wick product. These
terms can be paired with the corresponding terms
$$\sum_{\varphi,\chi_i}\pa_\mu\erw{T\varphi\,\chi_i}\cdot T\big(\wick{\frac{\pa Y}{\pa\varphi}\frac{\pa X_i}{\pa\chi_i}} \wick{X_1}\dots
\wick{\big\slash \!\!\!\!X_i}\dots \wick{X_n}\big)\tree$$
arising in the Wick expansion of of $\pa T\big({Y}
{X_1}\dots {X_n}\big)$, in which the derivative hits a contracted factor
of $Y$. Thus, by \eref{2pt-obs}, we have
$$[T,\pa_\mu]\big(\wick{Y}
\wick{X_1}\dots \wick{X_n}\big) = \sum_i
\sum_{\varphi,\chi_i}O_\mu(\varphi;\chi_i)\cdot T\Big(\wick{\frac{\pa Y}{\pa\varphi}\frac{\pa X_i}{\pa\chi}}\wick{X_1}\dots
\wick{\big\slash \!\!\!\!X_i}\dots \wick{X_n}\Big)\tree.$$
By \eref{OVexp}, this proves the claim. \qed

\newpage

{\em Proof of \pref{p:O3LV} (cont'd):} We need the case $n=2$ of
\lref{l:Lb}, where $Y^\mu=V_1^\mu$ is a vector field.  
With notation $O_{Y_1}(X',Z'')\equiv O_\mu(Y_1^\mu;X,Z)$, the first line of \eref{O3} can be written as
\bea{firstl}&&\SS_3\big(3O_{V_1}(K_1',K_1'') + 3O_{V_1}(\pa'V_1',K_1'') -
3\pa'O_{V_1}(V_1',K_1'')  +\notag \\ &&+ \, O_{V_1}(\pa' V_1',\pa''V_1'')-2\pa'O_{V_1}(V_1',\pa''V_1'')+\pa'\pa''O_{V_1}(V_1',V_1'')\big).\eea
By \lref{l:Lb}, and with some rearrangements, this becomes 
$$\SS_3\Big(3TO^{(2)}(x,x')K_1(x'') +
3O_{O_{V_1}(V_1')}(K_1'') + 
O_{V_1}\big(O_{V_1'}(3K_1''+2\pa''V_1'')\big)
-2\pa''O_{V_1}\big(O_{V_1'}(V_1'')\big)\Big)$$
with $O^{(2)}(x,x')$ as in \eref{O2O}.  
Expressing the latter by \eref{canc2} in terms of the second-order fields $L_2,K_2,V_2,
W_2$, one can cancel the term involving
$\delta_{x'x''}\cdot TK_1(L_2'-K_2'-\pa'V_2')$ in the last line of
\eref{O3}, and rewrite  all the remaining terms as \eref{O3O}. \qed

If one is interested only in the third-order densities $K_3$ and
$L_3$, it suffices to compute the integral over \eref{canc3} and demand:
\bea{ISI3}\int dx\,dx'\,dx''\,
O^{(3)}(x,x',x'') \stackrel!= \int dx\, (L_3(x)-K_3(x)).\eea
Because all total derivatives drop out, the
integral over \eref{O3O} reduces to
\bea{IO3O}\int \!\!dx\,dx'\,dx''\,
O^{(3)}(x,x',x'') = \int \!\!dx\,
\big(\Omega_{V_1}(3L_2-\Omega_{V_1}(K_1+2L_1)) +
3\Omega_{V_2-\Omega_{V_1}(V_1)}(K_1)\big)(x),\qquad\eea
where for vector fields $Y^\mu$,
$$\Omega_Y(X):= -i\int dx' \, O_Y(X') $$
coincides with \eref{Omdef} in the case when there are no derivatives of
$\delta$-functions. The form \eref{IO3O} is easy to evaluate, even when the two-point 
obstructions involve derivatives of $\delta$-functions. It then
suffices to equate the integrands of \eref{ISI3} and \eref{IO3O}.

\section{Propagators and two-point obstructions}
\label{a:prop}

We denote by $W_m(x-x')$ and $T_m(x-x')$ the canonical scalar
two-point function and time-ordered two-point function of mass $m$, such that
$iT_m$ is the Feynman propagator, and
$$(\square+m^2)W_m(x-x')=0, \qquad (\square+m^2)T_m(x-x')=-i\delta(x-x').$$

\subsection{Propagators and two-point obstructions for the Higgs field}
\label{a:higgs-2pt}
The two-point function of the Higgs field is the canonical scalar
two-point function of mass $m_H$:
$$\erw{H(x) H(x')}= W_{m_H}(x-x').$$
The two-point functions of derivatives of $H$ are derivatives
of $W_{m_H}$.
We define the ``kinematical'' propagators among the fields $H$ and $\pa H$ by the same differential operators acting on the massive Feynman propagator $iT_{m_H}(x-x')$. However, the
time-ordering prescription fixes the propagators only outside the
point $x-x'=0$. The freedom to add an arbitrary derivative of
$\delta(x-x')$ is constrained, apart from Lorentz covariance, by the
scaling degree that must not exceed the scaling degree of the
kinematical propagators. $T_{m_H}$ has the canonical scaling degree
2. Every derivative increases the scaling degree by 1. The
$\delta$-function has scaling degree 4. Therefore, $\erw{T\pa
  H\pa'H'}$ has a freedom of renormalization:
\bea{THH}\erw{T\pa_\mu
  H(x)\pa_\nu'H(x')} = -\pa_\mu\pa_\nu T_{m_H}(x-x')
+ic_H \eta_{\mu\nu}\delta(x-x')\eea
with an arbitrary real constant $c_H$.

Using $\square H=-m_H^2H$ 
and the definition \eref{2pt-obs}, one computes the two-point obstructions
$O_\mu(\varphi;\varphi')$ among the fields $H$ and $\pa H$, as
displayed in \eref{OH}:
$$O_\mu(H;\pa_\nu'H')=\erw {T\pa_\mu H\pa'_\nu H'}-\pa_\mu\erw {TH\pa'_\nu H'} = ic_H \eta_{\mu\nu}\delta(x-x'),$$
$$O_\mu(\pa^\mu H;H')=\erw {T\square HH'}-\pa_\mu\erw {T\pa^\mu HH'} =
-(m_H^2+\square)T_{m_H}(x-x') = i\delta(x-x'),$$
$$O_\mu(\pa^\mu H;\pa'_\nu H')=\erw {T\square H\pa'_\nu H'}-\pa_\mu\erw {T\pa^\mu H\pa'_\nu H'} =
-i(c_H +1)\pa_\nu\delta(x-x').$$

\subsection{Propagators and two-point obstructions for Krein space fields}
\label{a:krein-2pt}
The Feynman gauge two-point function of the Krein potential $A\K$ of
mass $m$ is
$$\erw{A\K_\mu(x) A\K_\nu(x')}=-\eta_{\mu\nu} W_m(x-x').$$
By the definitions
$\phi\K:= -m^{-2} (\pa A\K)$ and $B:= A\K-\pa\phi\K$ as in \sref{s:Krein}, one computes the
two-point functions (with Lorentz indices and arguments suppressed in
an obvious way)
$$\erw{\phi\K\phi\K{}'} = -m^{-2}W_m, \qquad
\erw{\phi\K\pa\phi\K{}'}=-\erw{\pa\phi\K\phi\K{}'}=\erw{\phi\K A\K{}'}=
-\erw{A\K\phi\K{}'}=m^{-2} \pa W_m,$$
$$\erw{\pa\phi\K\pa\phi\K{}'} = \erw{\pa\phi\K  A\K{}'}
=\erw{A\K\pa\phi\K{}'}=m^{-2} \pa\pa W_m,$$
$$\erw{BB'} =\erw{BA\K{}'} = \erw{A\K B'}=  -(\eta+m^{-2}\pa\pa)W_m,$$
$$\erw{B\phi\K{}'}=\erw{\phi\K B'}=\erw{B\pa\phi\K{}'}=\erw{\pa\phi\K B'}=0.$$
We define the kinematical propagators by the same differential
operators acting on the massive Feynman propagator $iT_m$. Only those
propagators involving $\pa_\mu\pa_\nu T_m$ have scaling degree 4 and admit a renormalization proportional to
$\eta_{\mu\nu}\delta(x-x')$. By linearity and 
$A\K_\mu-B_\mu=\pa_\mu\phi\K$, there is only one  independent parameter $c_B$:
$$\erw{TBB'}= \erw{TA\K B'}=\erw{TBA\K{}'} = -(\eta+m^{-2}\pa\pa)T_m(x-x') +ic_B\cdot m^{-2}\eta \delta(x-x'), $$
$$\erw{T\pa\phi\K\pa\phi\K{}'}=\erw{TA\K\pa\phi\K{}'} = \erw{T\pa\phi\K A\K{}'} = m^{-2} \pa\pa T_m(x-x') -i c_B\cdot m^{-2}\eta \delta(x-x').$$

With $\pa B=0$, $\pa A\K = -m^2\phi\K$, and $\square\phi\K =-m^2\phi\K$, one
computes the relevant two-point obstructions among the fields $A\K$,
$B$, $\phi\K$. The results are \eref{OBK}.

\subsection{Propagators and two-point obstructions for string-localized fields}
\label{a:sloc-2pt}

The positive-definite two-point function of the Proca field $B$ of
mass $m$ is
\bea{BB}\erw{B_\mu(x) B_\nu(x')}=-(\eta_{\mu\nu}+m^{-2}\pa_\mu\pa_\nu) W_m(x-x').\eea
We use the short-hand notation $(I_eX)(x):= \ioi ds\, X(x+se)$. By the definitions 
$\phi(e):= I_e(eB)$, $A(e):= B+\pa \phi(e) = B + I_e \pa (eB)$, one computes the
two-point functions
$$\erw{A_\mu B'_\nu} = -(\eta_{\mu\nu}+e_\nu I_e \pa_\mu)W_m, \qquad
\erw{B_\mu A'_\nu} = -(\eta_{\mu\nu}-e'_\mu I_{-e'} \pa_\nu)W_m, $$
$$\erw{A_\mu\phi'} = -(e'_\mu I_{-e'}+(ee')I_eI_{-e'}\pa_\mu) W_m,
\qquad
\erw{\phi A'_\nu} = -(e_\nu I_{e}-(ee')I_eI_{-e'}\pa_\nu) W_m, $$
$$\erw{A_\mu A'_\nu} = -(\eta_{\mu\nu}+ e_\nu I_e\pa_\mu - e'_\mu
I_{-e'}\pa_\nu - (ee')I_eI_{-e'}\pa_\mu\pa_\nu)W_m,$$
$$\erw{B_\mu \phi'} = -(e'_\mu I_{-e'}+ m^{-2}\pa_\mu)W_m, \qquad
\erw{\phi B'_\nu } = -(e_\nu I_{e}- m^{-2}\pa_\nu)W_m,$$
$$\erw{\phi\phi'} = -((ee')I_eI_{-e'} - m^{-2})W_m,$$
where it was used repeatedly that $\square W_m=-m^2 W_m$ and $(e\pa)(I_eX)(x)=-X(x)$.

We define
the kinematical propagators by the same differential and integral
operators as in the two-point functions, acting on
$iT_m(x-x')$. The propagator $i\erw{TBB}$ has scaling degree 4 and admits
the renormalization (the same as in the Krein space approach
\aref{a:krein-2pt})\footnote{The freedom of propagator renormalization
  in causal perturbation theory should not be confused with
  propagators in different gauges, like $R_\xi$ gauges. The latter
  would violate causality because of their momentum space denominators $k^2-\xi m^2$.}:
\bea{TBB}\erw{TB_\mu B_\nu'} =
-(\eta_{\mu\nu}+m^{-2}\pa_\mu\pa_\nu)T_m(x-x') +ic_B\cdot m^{-2}
\eta_{\mu\nu} \delta(x-x').\eea
By inspection of the scaling degree that is lowered by 1
by a string integration, one observes that all other propagators 
admit only renormalizations involving string-integrals over
$\delta$-functions.

The kinematical propagators produce
string-localized two-point obstructions $O_\mu(A^\mu;X')$, as displayed in \eref{OA}.
We have made a careful analysis of possible string-localized
renormalizations of the propagators. Their precise structure is
dictated by the scaling degree, 
Lorentz invariance, the number of
string integrations, homogeneity in $e$ and $e'$, and the identity
$-(e\pa)I_e=1$ which implies the axiality
property $e^\mu A_\mu=0$. It turns out that it
is impossible to make all string-localized two-point
obstructions vanish -- one would rather produce more of them.%
\footnote{
E.g., contributions to the obstruction $O_\mu(A^\mu;A_\nu')$ in
\eref{OA} could come from renormalizions of $\erw{TA^\mu A_\nu'}$ or
$\erw{T\phi A_\nu'}$ (via $\pa_\mu A^\mu = -m^2\phi$). But the
latter (scaling degree 1) admits no renormalization at all, and the renormalization of
the former (scaling degree 2) by $((ee')\eta_{\mu\nu}-e'_\mu e_\nu)
I_eI_{-e'}\delta(x-x')$ would (via $\pa_\mu \phi = A_\mu-B_\mu$)
 produce a non-zero obstruction
 $O_\mu(\phi;A'_\nu)$. }
There is thus the risk that the
second-order obstructions \eref{O2OLQ} and \eref{O2O} of the S-matrix 
become string-localized. In this case, they cannot be cancelled by admissible
higher-order densities, as outlined in \sref{s:PSI-imp}. In the
Abelian Higgs model, they do not occur, thanks to a characteristic
feature of the model (namely, the string-localized field $A$ does not
  appear within
$V_n$) that prefers the kinematical choice of
propagators for the string-localized fields, see \sref{s:ahm}. 

We take therefore all relevant propagators in the Proca sector except \eref{TBB} to be  the kinematical
ones. One can then directly compute the relevant two-point
obstructions.  One obtains \eref{OB} and \eref{OA}.

For the $L$-$Q$-approach in \sref{s:ahmLQ}, we also need obstructions $O_\mu(w;X')$
involving the field $w= \delta_c\phi(c)$. These obstructions vanish
because $O_\mu(\phi;X')$ in \eref{OB} and \eref{OA} are string-independent, and
$O_\mu(w;X')=\delta_c O_\mu(\phi;X')$.

In \sref{s:locobs}, we also need obstructions of the field strength
$G_{\mu\nu}=\pa_\mu B_\nu-\pa_\nu B_\mu$. The obstructions \eref{OG}
follow from the unique propagators 
\bea{TBG}\erw{TB^\kappa G_{\mu\nu}'}= (\delta^\kappa_\nu \pa_\mu
-\delta^\kappa_\mu \pa_\nu )T_m, \quad \erw{T\phi(e) G_{\mu\nu}'}=
(e_\nu \pa_\mu -e_\mu \pa_\nu )I_eT_m\eea
and the kinematical propagator
\bea{TAG} \erw{TA^\kappa(e) G_{\mu\nu}'}= \big((\delta^\kappa_\nu \pa_\mu
-\delta^\kappa_\mu \pa_\nu )+ (e_\nu \pa_\mu -e_\mu \pa_\nu
)\pa^\kappa I_e\big) T_m.\eea

\section{Useful identities}
\label{a:useful}
The following structures appear in the computation of $O^{(2)}$ as in
\eref{O2O}. Let throughout $X^{(\prime)}\equiv X(x^{(\prime)})$, 
$\pa^{(\prime)}\equiv\pa_{x^{(\prime)}}$, 
and $\delta_{xx'}\equiv\delta(x-x')$ and $\delta_{xx'x''}\equiv \delta_{xx'}\delta_{x'x''}$.

\blemma{l:2delta} It holds 
\bea{XdY} X\cdot\delta_{xx'}\cdot Y' + X'\cdot\delta_{xx'}\cdot Y &=& \delta_{xx'}\cdot 2XY, \\ \notag
X\cdot\pa_\alpha\delta_{xx'}\cdot Y' + X'\cdot\pa'_\alpha\delta_{xx'}\cdot Y &=& \delta_{xx'}\cdot X\lrpa_\alpha Y,\\ \notag
\pa'_\alpha(X\cdot\delta_{xx'}\cdot Y') + \pa_\alpha(X'\cdot\delta_{xx'}\cdot Y)
&=& \delta_{xx'}\cdot \pa_\alpha(XY),\\ \notag
\pa'_\alpha(X\cdot\pa_\beta\delta_{xx'}\cdot Y') + \pa_\alpha(X'\cdot\pa'_\beta\delta_{xx'}\cdot Y)
&=& (\pa'_\alpha\pa_\beta+\pa_\alpha\pa'_\beta)(\delta_{xx'}\cdot XY) -
\delta_{xx'}\cdot \pa_\alpha (Y\pa_\beta X).
\eea
\elemma
{\em Proof:}
The proof is elementary, using identities of the form $X\cdot \pa \delta_{xx'} \cdot Y' =
\pa(X\cdot \delta_{xx'} \cdot Y') - \pa X\cdot \delta_{xx'} \cdot Y'$, as well as $(\pa+\pa')\delta_{xx'}=0$.
\qed

The following structures appear in the computation of $O^{(3)}$ as in
\eref{O3simp}.

\blemma{l:3delta}
It holds  
\bea{XdY3}3\SS_3\big(\delta_{x'x''}\cdot \pa'
\delta_{xx'}\cdot X\big)&=&2\delta_{xx'x''}\cdot\pa X,\\ \notag
3\SS_3\big(\delta_{x'x''}\cdot \pa'
\delta_{xx'}\cdot XY'\big) &=& \delta_{xx'x''}\cdot (2Y\pa X-X\pa Y).\eea
\elemma
{\em Proof:} For the first identity, write
$$3\delta_{x'x''}\cdot \pa'
\delta_{xx'}\cdot X = -3\delta_{x'x''}\cdot \pa
\delta_{xx'}\cdot X = 3\delta_{xx'x''}\cdot \pa X - 3\pa
(\delta_{xx'x''}\cdot X).$$
In the second term,
$3\delta_{xx'x''}\cdot X = 
\delta_{xx'x''}\cdot (X+X'+X'')$ is separately symmetric. Apply the symmetrization:
$$3\SS_3\big(\delta_{x'x''}\cdot \pa'
\delta_{xx'}\cdot X) = 3\delta_{xx'x''}\cdot \pa X -
\frac13(\pa+\pa'+\pa'')\big(\delta_{xx'x''}\cdot (X+X'+X'')\big),$$
and use that $(\pa+\pa'+\pa'')\delta_{xx'x''}=0$ while
$\delta_{xx'x''}\cdot(\pa+\pa'+\pa'')(X+X'+X'')=
3\delta_{xx'x''}\cdot\pa X$. This proves the first identity. For the 
second identity write 
$$\pa'
\delta_{xx'}\cdot XY' = \pa'
(\delta_{xx'}\cdot XY') - 
\delta_{xx'}\cdot X\pa'Y' = \pa'
\delta_{xx'}\cdot XY - 
\delta_{xx'}\cdot X\pa Y, $$
and apply the first identity. \qed


\small
\begin{thebibliography}{99} \itemsep0mm
 
\bibitem{AS} A. Aste, G. Scharf:   Non-abelian gauge theories as a
    consequence of perturbative quantum gauge invariance. Int.\
    J. Mod.\ Phys.\ A14 (1999) 3421--3434.
    \bibitem{ADS} A. Aste, M. Dütsch, G. Scharf: On gauge invariance and
    spontaneous symmetry breaking. J. Phys.\ {A30} (1997) 5785--5792.
\bibitem{BD} F.-M. Boas, M. D\"utsch: The Master Ward Identity. Rev.\
  Math.\ Phys.\ 14 (2022) 977--1049.
  \bibitem{Bog}  N.N. Bogoliubov, D.V. Shirkov: Introduction to the Theory of Quantized Fields. Wiley, New
   York, NY, U.S.A. (1959).
\bibitem{Du} P. Duch: Massive QED, unpublished notes (2018).
\bibitem{Dt} M. Dütsch: From Classical Field Theory to Perturbative
Quantum Field Theory. Springer-Birkhäuser 2019.
   \bibitem{DGSV} M. Dütsch, J. Gracia-Bond\'{i}a, F. Scheck,
    J. V\'arilly: Quantum gauge models without (classical) Higgs
    mechanism. Eur.\ Phys.\ J. C69 (2010) 599--621.
  \bibitem{DKS} M. D\"utsch, F. Krahe and G. Scharf:
  ``Scalar QED Revisited'', Nuovo Cimento A106 (1993) 277--307.
\bibitem{DPR} M. D\"utsch, L. Peters, K.-H. Rehren: The Master
  Ward Identity for Scalar QED, Ann.\ H. Poinc.\ 22 (2021) 2893--2933.
  \bibitem{DS} M. Dütsch, G. Scharf: Perturbative gauge invariance:
  the electroweak theory. Ann. Phys. (Leipzig) 8 (1999) 359--387.
    \bibitem{EG} H. Epstein, V. Glaser: The role of locality in
  perturbation theory. Ann.\ Inst.\ H. Poinc.\ A19 (1973)  211--295.
\bibitem{FPS} R. Ferrari, L.E. Picasso, F. Strocchi: Some remarks on
  local operators in quantum electrodynamics, Commun.\ Math.\ Phys.\ 35 (1974) 25--38. 
\bibitem{FMS}J. Fr\"{o}hlich, G. Morchio, F. Strocchi: Charged sectors and
scattering states in quantum electrodynamics, Ann.\ Phys.\ 119 (1970) 241--284.
\bibitem{Fro} C. Fronsdal: Massless fields with integer spin, Phys.\ Rev.\ D18 (1978) 3624--3629.
  \bibitem{G1} C. Ga\ss: Renormalization in string-localized field theories:
a microlocal analysis,  Ann.\ H. Poinc.\  23 (2022) 3493--3523.
  \bibitem{G2} C. Ga\ss: Constructive aspects of string-localized
    quantum field theory, PhD thesis, Göttingen University (2022).
  \bibitem{GGM} C. Ga\ss, J. Gracia-Bond\'{i}a, J. Mund:
    Revisiting the Okubo–Marshak argument.  Symmetry 13 (2021) 1645.
\bibitem{GB} J. Gracia-Bond\'{i}a: The causal gauge
  principle. Contemp.\ Math.\ 539 (2011) 115--133.
  \bibitem{GMV} J. Gracia-Bond\'{i}a, J. Mund, J. V\'arilly: The chirality
    theorem, Ann.\ H. Poinc. 19 (2018) 843--874.
      \bibitem{GV} J. Gracia-Bond\'{i}a, J. V\'arilly: Ideas whose time
    has gone, arXiv:2207.06522.
\bibitem{Ja} A. Jaffe: High-energy behavior in quantum field
  theory. I. Strictly localizable fields. Phys.\ Rev.\ 158 (1967) 1454--1461.
  \bibitem{KO} T. Kugo, I. Ojima: Local covariant operator formalism
    of non-abelian gauge theories and quark confinement
    problem. Suppl.\ Prog.\ Theor.\ Phys.\ 66 (1979) 1--130. 
  \bibitem{MRS1} J. Mund, K.-H. Rehren,  B. Schroer: Helicity decoupling in the massless limit of massive
tensor fields. Nucl.\ Phys.\ B924 (2017) 699--727.
\bibitem{MRS2} J. Mund, K.-H. Rehren, B. Schroer: Gauss' Law and
string-localized quantum field theory. JHEP 01 (2020) 001.
\bibitem{MRS3} J. Mund, K.-H. Rehren, B. Schroer: Infraparticle
  fields and the formation of photon clouds. JHEP 04
  (2022) 083. 
\bibitem{MSY} J. Mund, B. Schroer, J. Yngvason: String-localized quantum
fields and modular localization. Commun.\ Math.\ Phys.\ 268 (2006) 621--672.
\bibitem{PS} M.E. Peskin, D.V. Schroeder: An Introduction to
  Quantum Field Theory, Pegasus Books, Reading (MA) 1995.
\bibitem{RR} H. Ruegg, M. Ruiz-Altaba: The Stueckelberg
  field. Int.\ J. Mod.\ Phys.\ A19 (2004) 3265--3347.
\bibitem{Scha} G. Scharf: 
    Quantum Gauge Theories: A True Ghost Story,
  Wiley (2001).
\bibitem{Sch11} B. Schroer: An alternative to the gauge theoretic
  setting. Found.\ of Phys.\  41 (2011) 1543--1568.
\bibitem{Sch19} B. Schroer: The role of positivity and causality in interactions
involving higher spin. Nucl.\ Phys.\ B941 (2019) 91--144.
\bibitem{Schw} M.D. Schwartz: Quantum Field Theory and the
    Standard Model, Cambridge University Press 2014.
\bibitem{Ti} F.~Tippner, ``Scalar QED with String-Localised
Potentials'', Bachelor's Thesis, G\"ottingen University, 2019.
\bibitem{Wein} S. Weinberg: The Quantum Theory of Fields, Cambridge University Press 1995. 
    \bibitem{Wig} E.P. Wigner: On unitary representations of the
  inhomogeneous Lorentz group. Ann.\  Math.\  (2nd Ser.)\ 40 (1939) 149--204. 
    \bibitem{Yng} J. Yngvason: Zero-mass infinite spin representations of the Poincar\'e group and quantum field
theory, Commun.\ Math.\ Phys.\ 18 (1970) 195
\bibitem{Wu} T.T. Wu, S.L Wu: Comparing the $R_\xi$ gauge and the unitary gauge for the
standard model: An example. Nucl.\ Phys.\ B914 (2017) 421–445.
  \end{thebibliography}
\end{document}